\documentclass{lmcs}
\pdfoutput=1

\usepackage{lastpage}
\lmcsdoi{17}{2}{17}
\lmcsheading{}{\pageref{LastPage}}{}{}%
{Jan.~30,~2019}{May~26,~2021}{}

\usepackage[utf8]{inputenc}
\usepackage{cmll}
\usepackage{amssymb}

\usepackage{tikz}
\usepackage{tikz-cd}
\usepackage{csquotes}
\usepackage{xspace}
\usepackage[normalem]{ ulem }
\usepackage{enumitem}
\usepackage[all]{xy}
\usepackage{xcolor}
 \definecolor{darkblue}{rgb}{0,0,0.6}
 \definecolor{darkgreen}{rgb}{0,0.45,0}
 \definecolor{darkred}{rgb}{0.75,0,0}
\usepackage{mathtools}
\mathtoolsset{centercolon}

\usepackage{xifthen}

\theoremstyle{definition}
\newtheorem{nonexample}[thm]{Non-example}

\newcommand{\UniMath}{\href{https://github.com/UniMath/UniMath}{\nolinkurl{UniMath}}\xspace}

\newcommand{\codebaseurl}{https://github.com/UniMath/largecatmodules}

\newcommand{\coqdocbasebaseurl}{https://initialsemantics.github.io/doc/cee7580}
\newcommand{\coqdocbaseurl}{\coqdocbasebaseurl /Modules.}
\newcommand{\urlhash}{\#}
\newcommand{\coqdocurl}[2]{\coqdocbaseurl #1.html\urlhash #2}
\newcommand{\nolinkcoqident}[1]{\nolinkurl{#1}} % TODO: give better def for this?
\makeatletter
\newcommand{\coqident}{\begingroup\@makeother\#\@coqident}
\newcommand{\@coqident}[3][]{%
  \ifthenelse{\isempty{#2}}%
  {\nolinkcoqident{#3}}%
  {\ifthenelse{\isempty{#1}}%
  {\href{\coqdocurl{#2}{#3}}{\nolinkcoqident{#3}}}%
  {\href{\coqdocurl{#2}{#3}}{\nolinkcoqident{#1}}}}%
\endgroup}
\newcommand{\gittagbaseurl}{https://github.com/UniMath/TypeTheory/tree/}
\newcommand{\gittagurl}[1]{\gittagbaseurl #1}
\newcommand{\gittag}{\begingroup\@makeother\#\@gittag}
\newcommand{\@gittag}[1]{\href{\gittagurl{#1}}{\nolinkurl{#1}}\endgroup}

\makeatother
\newcommand\Rquot{{\hat{\Upsilon}/F}}
\newcommand\LCb{\LC_{\beta\eta}}
\newcommand\epiar{epi-signature}
\newcommand{\maybeinv}{\ensuremath{\mathsf{Maybe}^{-1}}}
\newcommand{\pullback}{\ar@{}[dr]|<<{\ulcorner}}
\newcommand{\pushout}{\ar@{}[ul]|<<{\ulcorner}}
\newcommand{\coend}[1]{\ensuremath{\int^{p:\NN} {#1}^{(\underline{p})}\times {#1}^{\overline{p}}}}
\DeclareMathOperator{\Mod}{Mod}
\DeclareMathOperator{\Mon}{Mon}
\newcommand{\LMod}{\ensuremath{\int\Mod}}
\newcommand{\WRep}{\ensuremath{\int \Mon}}

\DeclareMathOperator{\colim}{colim}

\newcommand{\iso}{\simeq}
\newcommand{\rar}{\longrightarrow}
\newcommand{\LC}{{\mathsf{LC}}}
\newcommand{\var}{{\mathsf{var}}}
\newcommand{\app}{{\mathsf{app}}}
\newcommand{\abs}{{\mathsf{abs}}}

\newcommand{\fix}{{\mathsf{fix}}}

\newcommand{\Set}{{\mathsf{Set}}}
\newcommand{\subst}{{\mathsf{subst}}}

\newcommand{\Maybe}{{\mathsf{Maybe}}}

\newcommand{\Cont}{{\mathsf{Cont}}}
\newcommand{\NN}{{\mathbb{N}}}

\newcommand{\C}{\mathcal{C}}
\newcommand{\Sig}{\mathsf{Sig}}

\newcommand{\intsubst}{\mathsf{isubst}}

\newcommand{\fixcoend}[1]{\ensuremath{\int^{n:\NN}\underline{n}\times({#1}^{(\underline{n})})^{\overline{n}}}}

\newcommand{\listfunctor}{\ensuremath{\mathsf{list}}}
\newcommand{\squarefunctor}{\ensuremath{\mathsf{square}}}
\newcommand{\finpower}{\ensuremath{\PP_{\mathrm{fin}}}}
\newcommand{\PP}{\ensuremath{\mathcal{P}}}
\newcommand{\maybe}{\mathsf{Maybe}}
\newcommand{\inl}{\mathsf{inl}}
\newcommand{\inr}{\mathsf{inr}}

\newcommand{\symdsig}[1]{\mathcal{S}^{(#1)} \Theta}
\newcommand{\Signaturemap}[1]{\ensuremath{ {#1}^\sharp }}
\newcommand{\Signaturemapaction}[1]{\ensuremath{ \epsilon_{#1}}}
\newcommand{\Id}{\mathsf{Id}}

\newcommand{\SigStrength}{\mathsf{SigStrength}}

\begin{document}

\title{Presentable signatures and initial semantics}

\author[Benedikt Ahrens]{Benedikt Ahrens\rsuper{a}}
\address{\lsuper{a}University of Birmingham, United Kingdom}
\email{b.ahrens@cs.bham.ac.uk}
%{https://orcid.org/0000-0002-6786-4538}
%\thanks{Ahrens acknowledges the support of the Centre for Advanced Study (CAS) in Oslo, Norway, which funded and hosted the research project \emph{Homotopy Type Theory and Univalent Foundations} during the 2018/19 academic year.}

\author[Andr\'e Hirschowitz]{Andr\'e Hirschowitz\rsuper{b}}
\address{\lsuper{b}Université Côte d’Azur, CNRS, LJAD, France}
\email{ah@unice.fr}
%{https://orcid.org/0000-0003-2523-1481}{}

\author[Ambroise Lafont]{Ambroise Lafont\rsuper{c}}
\address{\lsuper{c}University of New South Wales, Sydney, Australia}
\email{a.lafont@unsw.edu.au}
%{https://orcid.org/0000-0002-9299-641X}{}

\author[Marco Maggesi]{Marco Maggesi\rsuper{d}}
\address{\lsuper{d}Università degli Studi di Firenze, Italy}
\email{marco.maggesi@unifi.it}
%{https://orcid.org/0000-0003-4380-7691}
%\thanks{Maggesi acknowledges the support of Gruppo Nazionale per le Struttore Algebriche, Geometriche e le loro Applicazioni (GNSAGA), Istituto Nazionale di Alta Matematematica ``F.~Severi'' (INdAM), and Italian Ministry of Education, University and Research (MIUR).}
%\authorrunning{B. Ahrens, A. Hirschowitz, A. Lafont, and M. Maggesi}

%\Copyright{Benedikt Ahrens, Andr\'e Hirschowitz, Marco Maggesi, Ambroise Lafont}%mandatory, please use full first names. LIPIcs license is "CC-BY";  http://creativecommons.org/licenses/by/3.0/

% \subjclass{Theory of computation $\rightarrow$ Formal languages and auto\-ma\-ta theory $\rightarrow$ Formalisms $\rightarrow$ Algebraic language theory}% mandatory: Please choose ACM 2012 classifications from https://www.acm.org/publications/class-2012 or https://dl.acm.org/ccs/ccs_flat.cfm . E.g., cite as "General and reference $\rightarrow$ General literature" or \ccsdesc[100]{General and reference~General literature}. 

%\ccsdesc[100]{Theory of computation~%Formal languages and auto\-ma\-ta theory~Formalisms~
%Algebraic language theory}

%mandatory

% \category{}%optional, e.g. invited paper

%\relatedversion{}%optional, e.g. full version hosted on arXiv, HAL, or other respository/website

%\supplement{Computer-checked proofs with compilation instructions on \url{\codebaseurl /tree/cee7580}}%optional, e.g. related research data, source code, ... hosted on a repository like zenodo, figshare, GitHub, ...

% \funding{This work has partly been funded by the CoqHoTT ERC Grant 637339
% and by the EPSRC Grant EP/T000252/1.
% This material is based upon work supported by the Air Force Office of
% Scientific Research under
% award number FA9550-17-1-0363.}%optional, to capture a funding statement, which applies to all authors. Please enter author specific funding statements as fifth argument of the \author macro.

%\acknowledgements{We thank the anonymous referees for their helpful and constructive comments.}%optional

\maketitle

\begin{abstract}
  We present a device for specifying and reasoning about
  syntax for datatypes, programming languages, and logic calculi. More
  precisely, we study a notion of \enquote{signature} for
  specifying syntactic constructions.

  In the spirit of Initial Semantics, we define the \enquote{syntax
    generated by a signature} to be the initial object---if it
  exists---in a suitable category of models.  In our framework, the
  existence of an associated syntax to a signature is not
  automatically guaranteed.  We identify, via the notion of
  \emph{presentation of a signature}, a large class of signatures that
  do generate a syntax.

  Our (presentable) signatures subsume classical algebraic signatures
  (i.e., signatures for languages with variable binding, such as the
  pure lambda calculus) and extend them to include several other
  significant examples of syntactic constructions.

  One key feature of our notions of signature, syntax, and
  presentation is that they are highly compositional, in the sense
  that complex examples can be obtained by gluing simpler ones.
  Moreover, through the Initial Semantics approach, our framework provides,
  beyond the desired algebra of terms, a well-behaved substitution and
  the induction and recursion principles associated to the syntax.

  This paper builds upon ideas from a previous attempt by
  Hirschowitz-Maggesi, which, in turn, was directly inspired by
  some earlier work of Ghani-Uustalu-Hamana and Matthes-Uustalu.

  The main results presented in the paper are computer-checked within
  the UniMath system.
\end{abstract}

\section{Introduction}

\subsection{Initial Semantics}

The concept of characterising data through an initiality property is standard in computer science, 
where it is known under the terms \emph{Initial Semantics} and \emph{Algebraic Specification} \cite{adj}, 
and has been popularised by the movement of \emph{Algebra of Programming} \cite{DBLP:books/daglib/0096998}.

This concept offers the following methodology to define a \emph{formal language}%
\footnote{Here, the word \enquote{language} encompasses data types, programming languages and logic calculi, as well as languages for algebraic structures as considered in Universal Algebra.}:
\begin{enumerate}
   \item Introduce a notion of signature.
   \item Construct an associated notion of model. 
         Such models should form a category.
   \item Define the \emph{syntax generated by a signature} to be its initial model, when it exists.
   \item Find a satisfactory sufficient condition for a signature to
     generate a syntax\footnote{In the literature, the word \enquote{signature}
       is often reserved for the case where such sufficient condition
       is automatically ensured.}.
\end{enumerate}
The models of a signature should be understood as domain of interpretation of the syntax generated by the signature:
initiality of the syntax should give rise to a convenient \emph{recursion} principle.

For a notion of signature to be satisfactory, it should satisfy the
following conditions:
\begin{itemize}
\item it should extend the notion of algebraic signature, and
\item complex signatures should be built by assembling simpler ones,
  thereby opening room for compositionality properties.
\end{itemize}

\noindent
In the present work, we consider a general notion of signature---together 
with its associated notion of model---which is suited for the specification
of untyped programming languages with variable binding.  On the one hand, our
signatures are fairly more general than those introduced in some of
the seminal papers on this topic
\cite{DBLP:conf/lics/FiorePT99,Harper:1993:FDL:138027.138060,DBLP:conf/lics/GabbayP99}, which are
essentially given by a family of lists of natural numbers indicating the number
of variables bound in each subterm of a syntactic construction (we
call them \enquote{algebraic signatures} below).  On the other hand,
the existence of an initial model in our setting is not automatically
guaranteed.

One main result of this paper is a sufficient condition on a signature
to ensure existence of an initial model.  Our condition is satisfied far
beyond the algebraic signatures mentioned above.  Specifically, our
signatures form a cocomplete category and our condition is preserved
by colimits (Section~\ref{s:rep-class}).  Examples are given in
Section \ref{s:examples}.

Our notions of signature and syntax enjoy modularity in the sense introduced by \cite{DBLP:journals/lisp/GhaniUH06}: 
indeed, we define a \enquote{total} category of models where objects are pairs consisting of a signature together with one of its models; 
and in this total category of models, merging two extensions of a syntax corresponds to building an amalgamated sum.

The present work improves on a previous attempt \cite{DBLP:journals/corr/abs-1202-3499} in two main ways: firstly, it gives a much simpler condition for the existence of an initial model; secondly, it provides computer-checked proofs for all the main statements.

\subsection{Computer-checked formalization}

This article is accompanied by computer-checked proofs of the main results (Theorem~\ref{thm:modularity}, Theorem~\ref{thm:pres_sig_repr}, and its variant, Theorem \ref{thm:is_right_adjoint}).
 These proofs are based on the \UniMath library \cite{UniMath},
 which itself is based on the proof assistant \textsf{Coq} \cite{Coq:manual}.
 The main reasons for our choice of proof assistant are twofold:
 firstly, the logical basis of the Coq proof assistant, dependent type theory, is
 well suited for abstract algebra, in particular, for category theory.
 Secondly, a suitable library of category theory, ready for use by us, 
 had already been developed \cite{DBLP:journals/mscs/AhrensKS15,DBLP:journals/lmcs/AhrensL19}.

 The formalization consists of
about 8,000 lines of code, and can be consulted on \url{\codebaseurl}.
A guide is given in the \href{\codebaseurl /blob/cee7580/README.md}{README}.

For the purpose of this article, we refer to a fixed version of our library,
with the short hash \href{https://github.com/UniMath/largecatmodules/tree/cee75804401f7f74b53900c997863e12065b0f7c}{cee7580}.
This version compiles with version \href{https://github.com/UniMath/UniMath/tree/bcc8344b5ce19ff1d8e915bf5ad338d25c9e4642}{bcc8344}
of \UniMath. 

Throughout the article, statements and human-readable proofs are accompanied by their corresponding identifiers in the formalization.
These identifiers are also hyperlinks to the online documentation stored at \url{\coqdocbasebaseurl /index.html}.

While a computer-checked proof does not constitute an absolute guarantee of correctness, it seems fair to affirm that it increases trustworthiness drastically.

\subsection{Related work}
\label{ss:related}

This paper is a counterpart to Chapter~3 of Lafont's PhD thesis
\cite{phd-lafont}.
The other chapters there treat related topics; they provide
\begin{itemize}
  \item
 a systematic approach to \emph{equations} for our notions of signature and
 model of a signature, see~\cite{DBLP:conf/rta/AhrensHLM19} and \cite[Chapter~4]{phd-lafont}; and
 \item
  two extensions of the present work accounting for operational
 semantics, see~\cite{AHLM:POPL2020} and \cite[Chapter~5]{phd-lafont}, and  \cite{DBLP:conf/fscd/HirschowitzHL20} and \cite[Chapter~6]{phd-lafont}, respectively.
\end{itemize}

The Initial Semantics approach to syntax has been introduced in the
seminal paper of Goguen, Thatcher, and Wagner~\cite{adj}.
The idea that the notion of monad is suited for modelling substitution concerning
syntax (and semantics) goes back at least to Bellegarde and Hook~\cite{DBLP:journals/scp/BellegardeH94} (see also, e.g., \cite{DBLP:journals/fac/BirdP99, Alt-Reus}).

In~\cite{DBLP:conf/lics/FiorePT99}, the authors propose a powerful theory which allows one to specify (by initiality) monoids in a monoidal category.
(Recall that monads are monoids in the category of endofunctors.)
They consider pointed strong endofunctors and cover binding signatures for finitary monads.

Fiore and collaborators studied equational theories on top of syntax with
variable binding.
In particular, they introduce the general notion of equational systems in~\cite{FHEqSys}. Moreover, given any
endofunctor with pointed strength, they construct an equational system whose algebras
are monoids equipped with an adequate action of this endofunctor.
Later, \cite{DBLP:conf/csl/FioreH10} tackles simply-typed syntax with equations,
relying under the hood on these equational systems.
Finally, \cite{DBLP:conf/mfcs/FioreM10} focuses on equational theories with variable
binding, in the style of Lawvere.

Pointed signatures with strength have been revisited by
Matthes and Uustalu~\cite{MU03} and Ghani, Uustalu, and Hamana~\cite{DBLP:journals/lisp/GhaniUH06} who consider a form of colimits (namely coends) of  such signatures.
Research on signatures with strength is
actively developed, see~\cite{monads_unimath} for a recent account.
In the preliminary version~\cite{DBLP:journals/corr/abs-1202-3499} of the present work, we also used pointed strengths. Here we take a slightly different path centered on the notion of module over a monad. In the present context, this
notion first appeared in the characterization by initiality
of the monad of the lambda calculus modulo $\alpha\beta\eta$-equivalence given by
Hirschowitz and Maggesi~\cite{Hirschowitz-Maggesi-2010}.
Any signature with strength gives rise to a signature in our sense, cf. Proposition~\ref{prop:sigs_w_strength}.

We should mention several other mathematical approaches to syntax and
semantics.

Following \cite{DBLP:journals/corr/AltenkirchCU14}, \emph{relative} monads and modules over them~\cite{ahrens_relmonads} allow one to incorporate operational semantics into the monadic view.
Accordingly, our present contribution could probably be customised for
this \enquote{relative} approach.

Another approach to syntax based on Lawvere Theories is illustrated in~\cite{HP07}, where Hyland and Power also outline the link with the language of monads and sketch some of the  history of the categorical research on syntax and semantics.

Gabbay and Pitts \cite{DBLP:conf/lics/GabbayP99} employ a different technique for 
modelling variable binding, based on nominal sets.
We do not see yet how our treatment of more general syntax carries
over to nominal techniques.

Finally, let us mention the classical approach based on Cartesian
closed categories recently revisited and extended by T.~Hirschowitz
\cite{DBLP:journals/corr/Hirschowitz14}.

\subsection{Organisation of the paper}

Section~\ref{s:modules} gives a succinct account of the notion of
module over a monad, which is the crucial tool underlying our
definition of signatures.
Our categories of signatures and models are described in Sections \ref{s:higher-order}
and \ref{s:actions}, respectively.
In Section~\ref{sec:syntax}, we give our definition of syntax, and we present our first main result, a modularity result about merging extensions of syntax.
Our notion of \emph{presentation of a signature} appears in Section \ref{s:rep-class}.
There, we also state our second main result: presentable signatures generate a syntax.	
The proof of that result is given in Section~\ref{sec:proof-theorem}.
In Section \ref{s:examples},  we give examples of presentable signatures.
In Section~\ref{sec:recursion}, we show through examples how recursion can be recovered from initiality.

\subsection{Publication history}
This is a revision of the conference paper \cite{ahrens_et_al:LIPIcs:2018:9671} presented at Computer Science Logic (CSL) 2018.
Besides several minor changes to improve overall readability, the following content has been added:
\begin{itemize}
 \item A comparison between signatures with strength and our signatures (Proposition~\ref{prop:sigs_w_strength}); 
 \item An analogue of Lambek's Lemma (Lemma~\ref{lem:lambek}),
       as well as an example of a signature that is not representable (Non-example~\ref{non-ex:powerset-not-repr});
 \item A variant of one of the main results (Theorem~\ref{thm:is_right_adjoint});
 \item A fix in Example~\ref{sec:recur-redexes} counting the redexes of a lambda term;
 \item A more uniform treatment of several examples---both new and previously presented---in Section~\ref{sec:post-comp-with-presentable-functor};
 \item Explicit statements about the use of the axiom of choice;
 \item Hyperlinks to an online documentation of the source code of our formalisation.
 \item We slightly changed the title of our work, so that it now mentions explicitly ``presentable signatures'', the main concept of the work.
\end{itemize}

\section{Categories of modules over monads}
\label{s:modules}

This work employs category theory extensively.
The essential notions required are those of category, functor, natural transformation, limit, adjunction, and monad.
Our primary reference on the subject is the classic handbook by Mac Lane~\cite{maclane}.

The main mathematical notion underlying our signatures is that of module over a monad.
In this section, we recall the definition and some basic facts about modules over
a monad in the specific case of the category $\Set$ of sets, although
most of this material is generalizable.
For a more extensive introduction to this topic we refer to  \cite{Hirschowitz-Maggesi-2010}.
Our running example is the untyped lambda calculus.

\subsection{Modules over monads}
\label{sec:modules-over-monads}

A \emph{monad} (over $\Set$) is a monoid in the
category $\Set\rar\Set$ of endofunctors of $\Set$, i.e., a triple
$R=(R,\mu,\eta)$ given by a functor $R\colon\Set \rar
\Set$, and two natural transformations $\mu\colon R\cdot R \rar R$ and
$\eta\colon I \rar R$ such that the following diagrams commute:
\begin{equation*}
%   \label{e:monad-axioms}
  \xymatrix@C=1.8cm@R=1.2cm{
    R \cdot R \cdot R \ar[r]^-{R\mu} \ar[d]_{\mu R} &
    R \cdot R \ar[d]^{\mu}\\
    R \cdot R \ar[r]_-{\mu} & M
  }
  \hskip1.2cm
  \xymatrix@C=1.8cm@R=1.2cm{
    R \cdot I \ar[r]^-{R\eta} \ar[rd]_{id} &
    R \cdot R \ar[d]_\mu &
    I \cdot R\ar[l]_-{\eta R}\ar[dl]^{id} \\
    {} & R & {}
  }
  % \mu \circ \mu R = \mu \circ R\mu, \qquad
  % \mu \circ \eta R = 1_R, \qquad
  % \mu \circ R \eta = 1_R \enspace .
\end{equation*}

\begin{exa}
 The functor $\Maybe : \Set \to \Set$ mapping a set $X$ to $X + 1$ is given a monad structure as follows.
 The unit $\eta : \Id \to \Maybe$ is given, on a set $X$, by the left inclusion $\inl : X \to X + 1$.
 The multiplication $\mu : \Maybe \cdot \Maybe \to \Maybe$ is given, on a set $X$,
 by the map $(X+1)+1 \to X + 1$ collapsing the two distinguished elements.
\end{exa}

\begin{exa}
 Our main example of monad is the monad of terms of the lambda calculus \cite{DBLP:journals/scp/BellegardeH94, Alt-Reus}. Its underlying functor $\LC : \Set \rar \Set$ is generated by three
 constructions,
 \begin{itemize}
  \item $\var : X \to \LC(X)$;
  \item $\app : \LC(X) \times \LC(X) \to \LC(X)$; and
  \item $\abs : \LC(X+1) \to \LC(X)$;
 \end{itemize}
 for any set $X$. 
 Here, the distinguished variable $\inr(*) \in X+1$ is the fresh variable that is bound by the construction $\abs$.
 The unit $\eta : \Id \to \LC$ is given by $\var$, and the multiplication $\mu : \LC \cdot \LC \to \LC$ is given by ``flattening'', which intuitively removes a layer of $\var$'s from within the terms in $\LC(\LC(X))$.
\end{exa}

Given two monads $R = (R, \eta, \mu)$ and $R' = (R',\eta',\mu')$, a \emph{morphism $f : R \rar R'$ of monads} 
is given by a natural transformation $f : R \rar R'$ between the underlying functors such that
the following diagrams commute:
\begin{equation*}
  \xymatrix@C=1.8cm@R=1.2cm{
    \Id \ar[r]^\eta \ar[dr]_\eta &
    R \ar[d]^-f \\
    {} & R' \\
  }
  \hskip1.2cm
  \xymatrix@C=1.2cm@R=1.2cm{
    R \cdot R \ar[r]^{f\,f}\ar[d]_\mu &
    R' \cdot R'\ar[d]^-{\mu'} \\
    R \ar[r]_f & R'
  }
  % f \circ \eta = \eta' , \qquad
  % f \circ \mu = \mu' \circ (f \cdot f) \enspace .
\end{equation*}

Now we want to explain in which sense syntactic constructions such as $\app$ and $\abs$ of $\LC$ commute with substitution. Indeed, it is \emph{not} the notion of morphism of monads that captures this commutativity:

\begin{nonexample}
 The natural transformation $\abs : \LC \cdot \Maybe \to \LC$ is \emph{not} a morphism of monads from the composition of monads $\LC \cdot \Maybe$ to $\LC$, see, e.g., \cite[Example 3.18]{DBLP:journals/jfrea/AhrensZ11}.
\end{nonexample}

Instead, we will explain how the natural transformation $\abs$ above is a morphism of \emph{modules over the monad $\LC$} \cite{HM}.

Let $R$ be a monad.

\begin{defi}[Modules]
  A \emph{(left) $R$-module} is given by a functor $M\colon \Set \rar \Set$
  equipped with a natural transformation $\rho^M\colon M \cdot R \rar
  M$, called \emph{module substitution}, which is compatible with the monad
  composition and identity, in the sense that the following two diagrams commute:
  \begin{equation*}
%     \label{e:module-axioms}
    \xymatrix@C=1.8cm@R=1.2cm{
      M \cdot R \cdot R \ar[r]^-{\rho^M R} \ar[d]_{M \rho^M} &
      M \cdot R \ar[d]^{\rho_M} \\
      M \cdot R \ar[r]_-{\rho_M} & M
    }
    \hskip1.2cm
    \xymatrix@C=1.8cm@R=1.2cm{
      M \cdot I \ar[r]^{M\eta} \ar[dr]_{id} & 
      M \cdot R \ar[d]^{\rho^M} \\
      {} & M
    }
    % \rho^M \circ \rho^M R = \rho^M \circ M \mu, \qquad
    % \rho^M \circ M\eta = 1_M.
  \end{equation*}
\end{defi}

There is an obvious corresponding definition of right $R$-modules that
we do not need to consider in this paper.  From now on, we will write
\enquote{$R$-module} instead of \enquote{left $R$-module} for brevity.

\begin{exas}\hfill
  \begin{enumerate}
    \item Every monad $R$ is a module over itself, which we call the
          \emph{tautological} module.
    \item For any functor $F\colon \Set \rar \Set$ and any $R$-module
          $M\colon \Set \rar \Set$, the composition $F\cdot M$ is an $R$-module
          (in the obvious way).
    \item   For every set $W$ we denote by $\underbar W \colon \Set
            \rar \Set$ the constant functor $\underbar W := X \mapsto W$.
            Then $\underbar W$ is trivially an $R$-module since $\underbar W =
            \underbar W \cdot R$.
          \item   Let $M_1$, $M_2$ be two $R$-modules.
            Then
            the product functor $M_1\times M_2$ is an $R$-module (see
            Proposition \ref{p:lim-colim-lmod} for a general
           statement).
        \item Let $R$ be a monad. Then $R \cdot \maybe$  is an $R$-module in a natural way (see Section~\ref{sec:deriv} for a general statement).
  \end{enumerate}
\end{exas}

\begin{defi}[Linearity]
  We say that a natural transformation of $R$-modules $\tau\colon M
  \rar N$ is
  \emph{linear}\footnote{
  Given a monoidal category $\C$, there is a notion of (left or right) module over a monoid object in $\C$ (see, e.g., \cite[Section~4.1]{brandenburg_phd} for details). The term \enquote{module} comes from the case of rings: indeed, a ring is just a monoid in the monoidal category of Abelian groups. Similarly, our monads are just the monoids in the monoidal category of endofunctors on $\Set$, and our modules are just modules over these monoids. Accordingly, the term \enquote{linear(ity)} for morphisms among modules comes from the paradigmatic case of rings.
  }
  if it is compatible with module substitution in $M$ and $N$:
  \begin{equation*}
    \xymatrix@C=1.8cm@R=1.2cm{
      M \cdot R \ar[r]^{\tau R} \ar[d]_{\rho^M} & 
      N\cdot R \ar[d]^{\rho^N} \\
      M \ar[r]_{\tau} & N
    }
    % \qquad
    % \tau \circ \rho^M = \rho^N \circ \tau R.
  \end{equation*}
  We take linear natural transformations as morphisms among modules.
  It can be easily verified that we
  obtain in this way a category that we denote $\Mod(R)$.
\end{defi}

\begin{exaC}[{\cite[Section~5.1]{HM}}]
 The natural transformations $\app : \LC \times \LC \to \LC$ and $\abs : \LC \cdot \Maybe \to \LC$ are morphisms of modules over the monad $\LC$.
\end{exaC}

Beyond binary products, the category $\Mod(R)$ of modules over some monad $R$ has general limits and colimits. These are constructed pointwise:

\begin{prop}[\coqident{Prelims.LModulesColims}{LModule_Colims_of_shape}, \coqident{Prelims.LModulesColims}{LModule_Lims_of_shape}]
  \label{p:lim-colim-lmod}
  $\Mod(R)$ is complete and cocomplete.
\end{prop}
These limits and colimits will in turn lift to signatures in Section~\ref{s:higher-order},
see Proposition~\ref{prop:signature-lim-colim}.

\subsection{The total category of modules}

We already introduced the category $\Mod(R)$ of modules with fixed
base $R$.  Here we consider a larger
category which collects modules with different bases.  To this end, we
need first to introduce the notion of pullback.
\begin{defi}[Pullback]\label{def:pb-mod}
  Let $f\colon R\rar S$ be a morphism of monads and $M$ an $S$-module.
  The composition
  \begin{math}
    M\cdot R \stackrel{Mf}\rar M \cdot S \stackrel{\rho^M}\rar M
  \end{math}
  turns the endofunctor $M$ into an $R$-module which is called \emph{pullback of $M$ along
  $f$} and denoted by $f^*\!M$.\footnote{The term \enquote{pullback} is standard in the terminology of Grothendieck fibrations (see Proposition~\ref{prop:bmod-fibration}).
  }
\end{defi}

\begin{defi}[Total module category]
  We define the \emph{total module category} $\int_{R}\Mod(R)$, or $\LMod$ for short, as follows\footnote{%
  Our notation for the total category is modelled after the category of elements
  of a presheaf, and, more generally, after the Grothendieck construction of a fibration.
  It overlaps with the notation for categorical ends.
  }:
 \begin{itemize}
  \item its objects are pairs $(R, M)$ of a monad $R$ and an $R$-module $M$.
  \item a morphism from $(R, M)$ to $(S, N)$ is a pair $(f, m)$ where
    $f\colon R \rar S$ is a morphism of monads, and $m\colon M \rar
    f^*N$ is a morphism of $R$-modules.  
 \end{itemize}
  The category $\LMod$ comes
    equipped with a forgetful functor to the category of monads, given
    by the projection $(R,M) \mapsto R$.
\end{defi}

\begin{prop}[\coqident{Prelims.LModulesFibration}{cleaving_bmod}]
  \label{prop:bmod-fibration}
 The forgetful functor $\LMod \to \Mon$ is a Grothendieck fibration
 with fibre $\Mod(R)$ over a monad $R$.
 In particular, any monad morphism $f : R \rar S$ gives rise to a functor
  \[f^*\colon \Mod(S) \rar \Mod(R)\]
 given on objects by Definition~\ref{def:pb-mod}.
\end{prop}

\begin{prop}[\coqident{Prelims.LModulesColims}{pb_LModule_colim_iso}, \coqident{Prelims.LModulesColims}{pb_LModule_lim_iso}]
For any monad mor\-phism
   $f : R \rar S$, the functor $f^* : \Mod(S) \rar \Mod(R)$ preserves limits and colimits.
\end{prop}

\subsection{Derivation}
\label{sec:deriv}

For our purposes, important examples of modules are given by the
following general construction. 
\begin{defi}[Derivation]
  For any $R$-module $M$, the \emph{derivative} of
  $M$ is the functor $M':= M\cdot \Maybe : X \mapsto M(X+1)$. It is an
  $R$-module with the substitution $\rho^{M'} \colon M'\cdot R\rar M'$ defined
  by the commutative diagram
  \begin{equation}
    \label{e:action-derived}
    \xymatrix@R=1.6cm@C=2cm{
      M(R(X)+*) \ar[r]^-{\rho^{M'}_X}
      \ar[d]_{M(R(i_X)+\eta_{X+*}\circ \underline*)} &
      M(X+*) \\
      M(R(X+*)) \ar[ru]_{\rho^M_{X+*}} & {}}
  \end{equation}
  where $i_X\colon X \rar X+*$ and $\underline *\colon * \rar X+*$ are
  the obvious maps.
\end{defi}
It is easy to check that derivation yields an endofunctor on the category $\Mod(R)$ of modules over a fixed monad $R$.
In particular, derivation can be iterated: we denote by $M^{(k)}$ the $k$-th derivative of $M$.
Moreover, notice that derivation preserves the product of modules (\coqident{Prelims.deriveadj}{commutes_binproduct_derivation}).

\begin{defi}
  \label{d:standard-modules}
  Given a list of nonnegative integers $\ell=[a_1,\dots,a_n]$ and a module
  $M$ over a monad $R$, we
  denote by $M^{\ell}=M^{[a_1,\dots,a_n]}$ the module
  $M^{(a_1)}\times\cdots\times M^{(a_n)}$.  Observe that, when
  $\ell=[]$ is the empty list, $M^{[]}$ is the final module $*$.
\end{defi}

\begin{defi}
  For every monad $R$ and $R$-module $M$ we define the \emph{(unary) substitution morphism}
\begin{math}
  \sigma \colon M'\times R \rar M
\end{math}
by $\sigma_X = \rho^M_X \circ w_X$,
where $w_X:M(X+*)\times R(X)\rightarrow M(R(X))$ is the map
\[
  w_X\colon (a,b) \mapsto M(\eta_X + \underline b)(a),\qquad
  \underline b \colon *\mapsto b .
\]
\end{defi}

\begin{lem}[\coqident{Prelims.deriveadj}{substitution_laws}]
  \label{l:linear-sigma}
  The transformation $\sigma$ is linear.
\end{lem}

The substitution $\sigma$ allows us to interpret the derivative $M'$
as the \enquote{module $M$ with one formal parameter added}.

  Abstracting over the module turns the substitution morphism into a
  natural transformation that is the unit of the following adjunction:

  \begin{prop}[\coqident{Prelims.deriveadj}{deriv_adj}]
  \label{t:derivation-adjunction}
  The endofunctor of $\Mod(R)$ mapping $M$ to the $R$-module $M\times R$
  is left adjoint to the derivation endofunctor, the unit being
  the substitution morphism~$\sigma$.
\end{prop}

\section{The category of signatures}
\label{s:higher-order}

In this section, we give our notion of signature.
 The destiny of a signature is to have actions
in monads.
An action of 
a signature $\Sigma$ in a monad $R$ should be a morphism from a module
$\Sigma(R)$ to the tautological one $R$.  For instance, in the case of the
signature $\Sigma$ of a binary operation, we have $\Sigma(R):= R^2 = R \times R$.
Hence a signature
assigns, to each monad $R$, a module over $R$ in a
functorial way.

\begin{defi}[\coqident{Signatures.Signature}{signature}]
  \label{def:signature}
  A \emph{signature} is a section of the forgetful functor
  from the category $\LMod$ to the category $\Mon$.
\end{defi}

Now we give our basic examples of signatures.

\begin{exas}
  \label{ex:signatures}
  \hfill
  \begin{enumerate}
  \item The assignment $R \mapsto R$ yields a signature, which we denote
    by $\Theta$.
    \label{ex:signatures:item:tautological}
  \item For any functor $F\colon \Set \rar \Set$ and any signature
    $\Sigma$, the assignment $R \mapsto F \cdot \Sigma(R)$ yields a
    signature which we denote $F \cdot \Sigma$.
    \label{ex:functor-signature}
  \item The assignment $R \mapsto *_R$, where $*_R$ denotes the final
    module over $R$, yields a signature which we denote by $*$.
  \item Given two signatures $\Sigma$ and $\Upsilon$, the assignment
    $R \mapsto \Sigma(R)\times \Upsilon(R)$ yields a signature which we
    denote by $\Sigma \times \Upsilon$.  For instance,
    $\Theta^2=\Theta\times\Theta$ is the signature of any
    (first-order) binary operation, and, more generally, $\Theta^n$ is
    the signature of $n$-ary operations.
  \item Given two signatures $\Sigma$ and $\Upsilon$, the assignment
    $R \mapsto \Sigma(R)+ \Upsilon(R)$ yields a signature which we denote
    by $\Sigma + \Upsilon$.  For instance, $\Theta^2+\Theta^2$ is the
    signature of a pair of binary operations.
  \end{enumerate}
\end{exas}

This last example explains why we do not need to distinguish here between 
\enquote{arities}---usually used to specify a single syntactic construction---and 
\enquote{signatures}---usually used to specify a family of syntactic constructions;
our signatures allow us to do both 
(via Proposition~\ref{prop:signature-lim-colim} for families 
that are not necessarily finitely indexed).

\emph{Elementary} signatures are of a particularly simple shape:
\begin{defi}% \label{def:elementary}
  For each list of nonnegative integers $\ell=[s_1,\dots,s_n]$, the
  assignment $R \mapsto R^{(s_1)}\times \cdots \times R^{(s_n)}$ (see
  Definition~\ref{d:standard-modules} and Example~\ref{ex:signatures}.\ref{ex:signatures:item:tautological}) is a signature, which we denote by
  $\Theta^{\ell}$, or by $\Theta'$ in the specific case of $s=[1]$.
  Signatures of this form are said
  \emph{elementary}.
\end{defi}

\begin{rem}
   \label{r:alg-ar-prod}
  The product of two elementary signatures is elementary.
\end{rem}

\begin{defi}[\coqident{Signatures.Signature}{signature_category}]
  \label{d:signature-morph}
  Given two signatures $\Sigma_1, \Sigma_2 \colon \Mon \rar \LMod$,
  a \emph{morphism of signatures from $\Sigma_1$ to $\Sigma_2$}
  is a natural transformation $m \colon \Sigma_1 \rar \Sigma_2$ which, post-composed
  with the projection $\LMod \rar \Mon$, becomes the
  identity.
  Signatures form a subcategory $\Sig$ of the category of functors from $\Mon$ to $\LMod$.
\end{defi}

  Limits and colimits of signatures can be easily constructed pointwise:
\begin{prop}[\coqident{Signatures.SignaturesColims}{Sig_Lims_of_shape},
                    \coqident{Signatures.SignaturesColims}{Sig_Colims_of_shape}, 
                    \coqident{Signatures.PresentableSignatureBinProdR}{Sig_isDistributive}]
  \label{prop:signature-lim-colim}
  \label{prop:signature-distributive}
  ~\\
  The category of signatures is complete and cocomplete.
  Furthermore, it is distributive: 
  for any signature $\Sigma$ and family of signatures $(S_o)_{o\in O}$,
  the canonical morphism
  \[\coprod_{o\in O} (S_o \times \Sigma)\rar
  (\coprod_{o\in O} S_o) \times \Sigma\] is an isomorphism.
\end{prop}

\begin{defi}
\label{def:algebraic}
  An \emph{algebraic signature} is a (possibly infinite) coproduct of elementary signatures.
\end{defi}
These signatures are those which appear in \cite{DBLP:conf/lics/FiorePT99}.
For instance, the algebraic signature of the lambda calculus is $\Sigma_\LC = \Theta^2 + \Theta'$.

To conclude this section, we explain the connection between 
\emph{signatures with strength} (on the category $\Set$) and our signatures.

Signatures with strength were introduced in \cite{MU03} (even though they were not given an explicit name there).
The relevant definitions regarding signatures with strength are summarized in \cite{monads_unimath}, 
to which we refer the interested reader.

We recall that a signature with strength \cite[Definition~4]{monads_unimath} is a pair of an endofunctor $H : [\C,\C] \to [\C,\C]$
together with a strength-like datum. Here, we only consider signatures with strength over the base category $\C:=\Set$.
Given a signature with strength $H$, we also refer to the underlying endofunctor on the functor category
$[\Set,\Set]$ as $H : [\Set,\Set] \to [\Set,\Set]$.

A morphism of signatures with strength \cite[Definition~5]{monads_unimath} is a natural transformation
between the underlying functors that is compatible with the strengths in a suitable sense.
Together with the obvious composition and identity, these objects and morphisms form a category $\SigStrength$ \cite{monads_unimath}.

 Any signature with strength $H$ gives rise to a signature $\tilde{H}$ \cite[Section~7]{DBLP:journals/corr/abs-1202-3499}.
 This signature associates, to a monad $R$, an $R$-module whose underlying functor is $H(UR)$,
 where $UR$ is the functor underlying the monad $R$.
 Similarly, given two signatures with strength $H_1$ and $H_2$, and a morphism 
 $\alpha : H_1 \to H_2$ of signatures with strength, we associate to it
 a morphism of signatures $\tilde{\alpha} : \tilde{H_1} \to \tilde{H_2}$.
 This morphism sends a monad $R$ to a module morphism 
 $\tilde{\alpha}(R) : \tilde{H_1}(R) \rar \tilde{H_2}(R)$
 whose underlying 
 natural transformation is given by $\alpha(UR)$, where, as before, $UR$ is the 
 functor underlying the monad $R$.
 These maps assemble into a functor:

\begin{prop}[\coqident{Signatures.SigWithStrengthToSignature}{sigWithStrength_to_sig_functor}]
\label{prop:sigs_w_strength}
 The maps sketched above yield a functor $\tilde{(-)} : \SigStrength \rar \Sig$. 
\end{prop}

\section{Categories of models}
\label{s:actions}

We define the notions of \emph{model of a signature} and \emph{action of a signature
in a monad}.

\begin{defi}[Actions and models]
  Let $\Sigma$ be a signature.
  Given a monad $R$,
  an \emph{action of $\Sigma$ in $R$} is an $R$-module morphism
  $r : \Sigma(R) \to R$.
  A \emph{model of $\Sigma$} is a pair $(R,r)$
  of a monad $R$ equipped with an action of
  $\Sigma$ in $R$.%
  \footnote{This terminology is borrowed from the vocabulary of
    algebras for a functor: an algebra for an endofunctor $F$ on a category
    $\C$ is an object $X$ of $\C$ with a morphism $\nu : F(X) \rar X$.
    This
    morphism is sometimes called an action.}
  A \emph{morphism of models of $\Sigma$ from $(R, r)$ to $(S, s)$} is a morphism of monads
  $m : R \to S$ compatible with the actions, in the sense that the
  following diagram of $R$-modules commutes:
  \[
    \xymatrix@C=1.6cm@R=1.2cm{
    %\xymatrix{
      **[l] \Sigma(R) \ar[r]^{r}\ar[d]_{\Sigma(m)} & **[r] R \ar[d]^m \\
      **[l] m^* (\Sigma(S)) \ar[r]_{m^*s} & **[r] m^* S}
  \]
  Here, the horizontal arrows come from the actions, the left vertical
  arrow comes from the functoriality of signatures, and
  $m\colon R \rar m^* S$ is the morphism of monads seen as morphism of
  $R$-modules.
\end{defi}

\begin{exa}
  The usual $\app\colon \LC^2 \rar \LC$ is an action of the elementary signature
  $\Theta^2$ in the monad $\LC$ of syntactic lambda calculus.
  The usual $\abs\colon \LC' \rar \LC$ is an action of the elementary signature
  $\Theta'$ in the monad $\LC$. 
  Then $[\app, \abs] : \LC^2 + \LC' \rar \LC$ is an action of the algebraic signature
  of the lambda calculus $\Theta^2 + \Theta'$ in the monad $\LC$.
\end{exa}

\begin{prop}
  \label{prop:cat-of-models}
  Let $\Sigma$ be a signature. Models of $\Sigma$, and their morphisms,
  together with the obvious composition and identity, form a category.
\end{prop}
\begin{defi}
  We denote the category of Proposition~\ref{prop:cat-of-models} by $\Mon^\Sigma$. 
  It comes equipped with a forgetful
  functor to the category of monads.
\end{defi}
In the formalisation, this category is recovered as the fiber category over $\Sigma$
of the displayed category \cite{DBLP:journals/lmcs/AhrensL19} of models, 
see \coqident{Signatures.Signature}{rep_disp}.
We have also formalized a direct definition (\coqident{Signatures.ModelCat}{rep_fiber_category})
and shown that the two definitions yield isomorphic categories: \coqident{Signatures.ModelCat}{catiso_modelcat}.

The following notion will be useful in the next section in Lemma~\ref{lem:models-fibration}.
\begin{defi}[Pullback]\label{def:pb-model}
  Let $f\colon \Upsilon \rar \Sigma $ be a morphism of signatures
     and $(R,r)$ a model of $\Sigma$.
  The linear morphism
  \begin{math}
    \Upsilon(R) \stackrel{f(R)}\rar \Sigma(R)\stackrel{r}\rar R
  \end{math}
  defines an action of $\Upsilon$ in $R$. The induced model of $\Upsilon$ is called
  \emph{pullback of $(R,r)$ along $f$} and denoted by $f^*\!(R,r)$.
\end{defi}

\section{Syntax}\label{sec:syntax}

We are primarily interested in the existence of an initial object in
the category $\Mon^\Sigma$ of models of a signature $\Sigma$.
We call such an essentially unique object \emph{the syntax generated by~$\Sigma$}.

\subsection{Representations of a signature}
\label{ss:representability}

\begin{defi}
  \label{d:representable}
  
  If $\Mon^{\Sigma}$ has an initial object, this object is essentially unique; 
  we say that it is a \emph{representation of $\Sigma$} and call it the 
  \emph{syntax generated by $\Sigma$}, denoted by $\hat{\Sigma}$.
  By abuse of notation, we also denote by $\hat{\Sigma}$ the monad underlying the model $\hat{\Sigma}$.
  
  If an initial model for $\Sigma$ exists, we say that $\Sigma$ is \emph{representable}%
  \footnote{For an algebraic signature $\Sigma$ without binding constructions, the map assigning to
    any monad $R$ its set of $\Sigma$-actions can be upgraded into a functor which is corepresented by
    the initial model.
  }.
\end{defi}

In this work, we aim to identify signatures that are representable.
This is not automatic: below, in Non-example~\ref{non-ex:powerset-not-repr}, we give a signature that is not representable.
Afterwards, we give suitable sufficient criteria for signatures to be representable.

We deduce the counter-example as a simple consequence of
a stronger result that we consider interesting in itself: an analogue of Lambek's Lemma \cite{Lambek1968}, 
given in Lemma~\ref{lem:lambek}.

The following preparatory lemma explains how to construct a new model
of a signature $\Sigma$ from a given one:

\begin{lem}\label{lem:mon_id_action}
  Let $(R,r)$ be a model of a signature $\Sigma$.
  Let $\eta:\mathsf{Id}\to R$ be the unit of the monad $R$, and
    let $\rho^{\Sigma(R)}:\Sigma(R)\cdot R\to\Sigma(R)$ be the module substitution
    of the $R$-module $\Sigma(R)$.
  \begin{enumerate}
    \item
  The injection
  $\mathsf{Id} \to \Sigma(R)+\mathsf{Id}$ together with
  the natural transformation
  \[
  \xymatrix@R=1.2cm{
    (\Sigma(R) + \mathsf{Id})\cdot (\Sigma(R) + \mathsf{Id})
    \iso
    \Sigma(R)\cdot (\Sigma(R) + \mathsf{Id}) + \Sigma(R) + \mathsf{Id}
    \ar[d]^{\Sigma(R)[r,\eta]+\_+\_}
    \\
    \Sigma(R)\cdot R + \Sigma(R) + \mathsf{Id}
    \ar[d]^{[\rho^{\Sigma(R)},id]+\_}
    \\
    \Sigma(R) + \mathsf{Id}
    }
  \]
  give the endofunctor $\Sigma(R) + \mathsf{Id}$ the structure of a monad.
  \item Moreover, this monad can be given the following $\Sigma$-action:
  \begin{equation}\label{eq:mon_id_action}
  \xymatrix@C=5em{
    \Sigma\bigl(\Sigma(R) + \mathsf{Id}\bigr)
    \ar[r]^-{\Sigma([r,\eta])}
    &
    \Sigma(R)\cdot R 
    \ar[r]^-{\rho^{\Sigma(R)}}
    &
    \Sigma(R)
    \ar[r]
    &
    \Sigma(R) + \mathsf{Id}
    }
  \end{equation}
  \item The natural transformation $[r,\eta]:\Sigma(R)+\mathsf{Id}\to R$ is a model morphism,
        that is, it commutes suitably with the $\Sigma$-actions of Diagram~\eqref{eq:mon_id_action}
        in the source and $r : \Sigma(R) \rar R$ in the target.
    \end{enumerate}
\end{lem}
In the computer-checked library, the construction of the model and of the model morphism are given in \coqident{Signatures.ModelCat}{mod_id_model} and
\coqident{Signatures.ModelCat}{mod_id_model_mor}, respectively.

\begin{defi}
  Given a model $M$ of $\Sigma$, we denote by $\Signaturemap{M}$ 
  the $\Sigma$-model constructed in Lemma~\ref{lem:mon_id_action},
  and by $\Signaturemapaction{M} : \Signaturemap{M} \rar M$
  the morphism of models defined there.
\end{defi}

\begin{lem}[\coqident{Signatures.ModelCat}{iso_mod_id_model}]
  \label{lem:lambek}
  If $\Sigma$ is representable, then the morphism of $\Sigma$-models
  \[ \Signaturemapaction{\hat{\Sigma}} : \Signaturemap{\hat{\Sigma}}  \rar \hat{\Sigma} \]
  is an isomorphism.
\end{lem}

\noindent
Now, we are able to give a non-representable signature.
\begin{nonexample}
  \label{non-ex:powerset-not-repr}
  
  Let $\PP$ denote the powerset functor and consider 
  the signature $\PP \cdot \Theta$ (see Example~\ref{ex:signatures}, Item~\ref{ex:functor-signature}): it
  associates, to any monad $R$, the module $\PP \cdot R$ that sends a set $X$ to the powerset $\PP(RX)$
  of $RX$. 
  This signature is not representable, otherwise
from Lemma~\ref{lem:lambek} we would have $\PP\hat{\Sigma}X + X \cong \hat{\Sigma}X$.
In particular,
we would have an injective map from $\PP\hat{\Sigma}X$ to $\hat{\Sigma}X$---%
contradiction.
 
\end{nonexample}

On the other hand, as a starting point, we can identify the following class of representable signatures:

\begin{thm}[\coqident{Signatures.BindingSig}{algebraic_sig_representable}]
  \label{t:alg-sig-representable}
  Algebraic signatures are representable.
\end{thm}

This result is proved
in a previous work \cite[Theorems 1 and 2]{HM}.
The construction of the syntax proceeds as follows:
an algebraic signature induces an endofunctor on the category of endofunctors on $\Set$. 
Its initial algebra (constructed as the colimit of the initial chain) is given the structure of a monad with an action of the algebraic signature, 
and then a routine verification shows that it is actually initial in the category of models.
The computer-checked proof 
uses the construction of a monad from an algebraic signature formalized in \cite{monads_unimath}.

In Section~\ref{s:rep-class}, we show a more general representability result:
Theorem~\ref{thm:pres_sig_repr} states that \emph{presentable} signatures, 
which form a superclass of algebraic  signatures, are representable.

\subsection{Modularity}
\label{ss:modularity}

In this section, we study the problem of how to merge two syntax extensions.
Our answer, a \enquote{modularity} result (Theorem~\ref{thm:modularity}), was stated already in 
the preliminary version \cite[Section~6]{DBLP:journals/corr/abs-1202-3499}, there without proof.

Suppose that we have a pushout square of representable signatures,
\[
  \xymatrix{
    \Sigma_0\ar[r]\ar[d] & \Sigma_1\ar[d]\\
    \Sigma_2\ar[r] & \Sigma \pushout
    }
\]
Intuitively, the signatures $\Sigma_1$ and $\Sigma_2$ specify two extensions of 
the signature $\Sigma_0$, and $\Sigma$ is the smallest extension containing both these extensions.
Modularity means that the corresponding
diagram of representations,
\[
  \xymatrix{
    \hat \Sigma_0\ar[r]\ar[d] & \hat\Sigma_1\ar[d]\\
    \hat \Sigma_2\ar[r] & \hat\Sigma}
\]
is a pushout as well---but we have to take care to state this in the \enquote{right} category.
The right category for this purpose is the following:

\begin{defi}[Total category of models]
We denote by $\int_{\Sigma} \Mon^{\Sigma}$, or $\WRep$ for short, the \emph{total category of models}:
\begin{itemize}
\item An object of $\WRep$ is a triple $(\Sigma, R, r)$ where $\Sigma$ is a signature, 
 $R$ is a monad, and $r$ is an action of $\Sigma$
  in $R$.
\item A morphism in $\WRep$ from $(\Sigma_1, R_1, r_1)$ to $(\Sigma_2, R_2, r_2)$ 
  consists of a pair $(i,m)$ of a signature morphism $i: \Sigma_1 \rar \Sigma_2$
  and a morphism $m$ of $\Sigma_1$-models from $(R_1, r_1)$
  to $(R_2, i^*(r_2))$.
\item It is easily checked that the obvious composition turns $\WRep$
  into a category.
\end{itemize}
\end{defi}

\begin{lem}[\coqident{Signatures.Signature}{rep_cleaving}]
  \label{lem:models-fibration}
  The projection $\pi : \WRep \to \Sig$ is a Grothendieck fibration.
  In particular, given a morphism $f\colon \Upsilon \rar \Sigma $ of signatures,
  the pullback map 
  defined in Definition~\ref{def:pb-model}
  extends to a functor
  \[ f^* : \Mon^{\Sigma} \rar \Mon^{\Upsilon} \enspace . \]
\end{lem}

Now for each signature $\Sigma$, we have an obvious inclusion from
the fiber $\Mon^\Sigma$ into $\WRep$, through which we may see 
the syntax $\hat \Sigma$ of any representable signature as an
object in $\WRep$.  Furthermore, a morphism $i\colon \Sigma_1 \rar
\Sigma_2$ of representable signatures yields a morphism $i_*:= \hat \Sigma_1 \rar
\hat \Sigma_2$ in $\WRep$.  Hence our pushout square of
representable signatures as described above yields a square in $\WRep$.

\begin{thm}[\coqident{Signatures.Modularity}{pushout_in_big_rep}]
\label{thm:modularity}
  Modularity holds in $\WRep$, in the sense that given a pushout
  square of representable signatures as above, the
  associated square in $\WRep$ is a pushout again.
\end{thm}

\begin{rem}
Note that Theorem~\ref{thm:modularity} does \emph{not} say that a pushout of representable
signatures is representable again; it only tells us that if all of the signatures in a pushout square are 
representable, then the syntax generated by the pushout is the pushout of the syntaxes.
In general, we do not know whether a colimit (or even a binary coproduct) of representable signatures is representable again.
\end{rem}

\section{Presentations of signatures and syntaxes}
\label{s:rep-class}

In this section, we identify a superclass of algebraic signatures that
are still representable: we call them \emph{presentable} signatures.

\begin{defi}[\coqident{Signatures.PresentableSignature}{isPresentable}]
  \label{d:presentable--sig}  
   Given a signature $\Sigma$, a \emph{presentation%
            \footnote{In algebra, a presentation of a group $G$ is an
                      epimorphism $F \to G$ where $F$ is free (together with a generating set of
                      relations among the generators).}
   of $\Sigma$}
   is given by an algebraic signature $\Upsilon$ and an epimorphism
   of signatures $p : \Upsilon \rar \Sigma$. %
  In that case, we say that \emph{$\Sigma$ is presented by $p : \Upsilon \rar \Sigma$}.
  A signature for which a presentation exists is called \emph{presentable}.
   \end{defi}

Of course, any algebraic signature is presentable.
   
Unlike representations, presentations for a signature are not essentially unique;
indeed, signatures can have many different presentations.

\begin{rem}
  By definition, any construction which can be encoded through a presentable signature $\Sigma$
  can alternatively be encoded through any algebraic signature \enquote{presenting} $\Sigma$.
  The former encoding is finer than the latter in the sense that terms
  which are different in the latter encoding can be identified by the
  former. In other words, a certain amount of semantics is integrated
  into the syntax.
\end{rem}

The main desired property of our presentable signatures is that, 
thanks to the following theorem, they are representable:

\begin{thm}[\coqident{Signatures.PresentableSignature}{PresentableisRepresentable}]
\label{thm:pres_sig_repr}
 Any presentable signature is representable.
\end{thm}
The proof is discussed in Section~\ref{sec:proof-theorem}.

Using the axiom of choice, we can prove a stronger statement:
\begin{thm}[\coqident{Signatures.EpiSigRepresentability}{is_right_adjoint_functor_of_reps_from_pw_epi_choice}]
 \label{thm:is_right_adjoint}
 We assume the axiom of choice.
 Let $\Sigma$ be a signature, and let $p : \Upsilon \rar \Sigma$ be a presentation of $\Sigma$.
 Then the functor $p^* : \Mon^{\Sigma}\rar\Mon^{\Upsilon}$
 has a left adjoint.
\end{thm}

In the proof of Theorem~\ref{thm:is_right_adjoint}, the axiom of choice is used to show that
endofunctors on $\Set$ preserve epimorphisms.

Theorem~\ref{thm:pres_sig_repr} follows from Theorem~\ref{thm:is_right_adjoint} 
since the left adjoint $p^{!} : \Mon^{\Upsilon}\rar\Mon^{\Sigma}$ preserves
colimits, in particular, initial objects.
However, our (formalized) proof of Theorem~\ref{thm:pres_sig_repr} discussed in Section~\ref{sec:proof-theorem} does not invoke the axiom of choice:
there, only some specific endofunctor on $\Set$
is considered, for which preservation of epimorphisms can be proved without 
using the axiom of choice.

For the examples of Section~\ref{s:examples}, we will use the following constructions of presentable signatures:

\begin{prop}[\coqident{Signatures.PresentableSignatureBinProdR}{har_binprodR_isPresentable}]
  Given a presentable signature $\Sigma$, the product signature $\Sigma \times \Theta$
  of $\Sigma$ and the tautological signature is again presentable.
\end{prop}
More generally, if $\Sigma_1$ and
  $\Sigma_2$ are presented by $\coprod_i\Upsilon_i$ and
  $\coprod_j\Phi_j$ respectively, then $\Sigma_1 \times \Sigma_2$ is
  presented by $\coprod_{i,j}\Upsilon_i \times \Phi_j$.

\begin{prop}\label{prop:colim-of-presentables}
 Any colimit of presentable signatures is presentable.
\end{prop}

\begin{cor}\label{cor:colim-presentable}
 Any colimit of algebraic signatures is representable.
\end{cor}

\section{Proof of Theorem~\ref{thm:pres_sig_repr}}
\label{sec:proof-theorem}

In this section, we prove Theorem~\ref{thm:pres_sig_repr}.
This proof is mechanically checked in our library; the reader
may thus prefer to look at the formalised statements in the library.

Note that the proof of Theorem~\ref{thm:pres_sig_repr} rests on the more technical Lemma~\ref{lem:th-epi-ini} below.

We will need the following characterization of epimorphisms of signatures.
\begin{prop}[\coqident{Signatures.EpiArePointwise}{epiSig_equiv_pwEpi_SET}]
\label{t:signature-epi-morph}
{Epimorphisms of signatures are exactly pointwise epimorphisms.}
\end{prop}
\begin{proof}
In any category, a morphism $f:a\rightarrow b$ is an epimorphism if and only if
the following diagram is a pushout diagram (\cite[Exercise III.4.4]{maclane}) :
\[
\begin{tikzcd}[sep=large]
a
\arrow[r, "f"] \arrow[d, "f",swap]  & b
\arrow[d, "id"]
\\
b  \arrow[r, "id",swap] & b
\end{tikzcd}
\]
Using this characterization of epimorphisms, the proof follows from the fact that colimits are computed pointwise in the category of signatures.
\end{proof}

Another important ingredient will be the following quotient construction for monads.
Let $R$ be a monad preserving epimorphisms, and let $\sim$ be a \enquote{compatible} family of relations on (the functor underlying) $R$, 
that is, for any $X : \Set_0$, $\sim_X$ is an equivalence relation on $RX$
 such that, for any $f : X \to Y$, the function $R(f)$ maps related
 elements in $RX$ to related elements in $RY$.
 Taking the pointwise quotient, we obtain a quotient $\pi:R \to \overline{R}$ in the functor category, 
 satisfying the usual universal property.
 We want to equip $\overline{R}$ with a monad structure that upgrades $\pi:R \to \overline{R}$ into a quotient in 
 the category of monads.
 In particular, this means that we need to fill in the square
 \[
   \begin{xy}
     \xymatrix@C=5em@R=1.2cm{
                        **[l] R\cdot R \ar[d]_{\pi\cdot \pi} \ar[r]^{\mu} &   R  \ar[d]^{\pi}
                   \\
                        **[l] \overline{R}  \cdot \overline{R} \ar@{-->}[r]_{\overline{\mu}} & \overline{R}
              }
   \end{xy}
 \]
 with a suitable $\overline{\mu} : \overline{R}  \cdot \overline{R} \rar \overline{R}$ satisfying the monad laws. But since $\pi$, and hence $\pi \cdot \pi$,
 is epi as $R$ preserves epimorphisms, this is possible when any two
 elements in $RRX$ that are mapped to the same element by $\pi \cdot \pi$ (the left vertical morphism) are also mapped to the same element by
 $\pi \circ \mu$ (the top-right composition).
 It turns out that this is the only extra condition needed for the upgrade.
 We summarize the construction in the following lemma:
 \begin{lem}[\coqident{Prelims.quotientmonad}{projR_monad}]
   \label{lem:quotient-monad}
   Given a monad $R$ preserving epimorphisms, and a compatible relation $\sim$ on $R$ such that
   for any set $X$ and $x,y \in RRX$, we have that
   if $(\pi \cdot \pi)_{X}(x) \sim (\pi\cdot \pi)_{X}(y)$ then $\pi(\mu(x)) \sim \pi(\mu(y))$.
   Then we can construct the quotient $\pi : R \to \overline{R}$ in the category of monads,
   satisfying the usual universal property.
 \end{lem}

\begin{defi}\label{def:epi-signature}
  An \emph{\epiar} is a signature $\Sigma$ that preserves the
  epimorphicity in the category of endofunctors on Set: for any monad
  morphism $f : R \rar S$, if $U(f)$ is an epi of functors, then so is
  $U(\Sigma(f))$.  Here, we denote by $U$ the forgetful functor from
  monads resp.~modules to endofunctors.
\end{defi}

If we admit the axiom of choice, then epimorphisms in $\Set$ have a retraction,
and thus any endofunctor on $\Set$ preserves epimorphisms.
Hence, in that case, any signature is an epi-signature, and the previous definition becomes superfluous.

\begin{exa}[\coqident{Signatures.BindingSig}{BindingSigAreEpiSig}]
  \label{ex:alg-epi-ar}
  Any algebraic signature is an \epiar{}.
\end{exa}

We are now in a position to state and prove the main technical lemma:

\begin{lem}[\coqident{Signatures.EpiSigRepresentability}{push_initiality}]
\label{lem:th-epi-ini}
Let $\Upsilon$ be representable, such that both $\hat{\Upsilon}$ and $\Upsilon(\hat{\Upsilon})$ preserve epimorphisms (as noted above, this condition is automatically fulfilled if one assumes the axiom of choice).
Let $F:\Upsilon \rightarrow \Sigma$ be a morphism of signatures.
Suppose that $\Upsilon$ is an \epiar{} and $F$ is an epimorphism.
Then $\Sigma$ is representable.
\end{lem}

\begin{proof}[Sketch of the proof]
As before, we denote by $\hat{\Upsilon}$ the initial $\Upsilon$-model, as well as---by abuse of notation---%
its underlying monad.
For each set $X$,
we consider the equivalence relation $\sim_X$ on $\hat{\Upsilon}(X)$ defined as
follows: for all $x, y \in \hat{\Upsilon}(X)$ we stipulate that $x \sim_X y$ if and
only if $i_X (x) = i_X (y)$ for each (initial) morphism of
$\Upsilon$-models $i : \hat{\Upsilon} \rightarrow F^* S$ with $S$ a
$\Sigma$-model and $F^* S$ the $\Upsilon$-model induced by
$F:\Upsilon\rightarrow \Sigma$.

By virtue of Lemma~\ref{lem:quotient-monad}, since $\hat{\Upsilon}$ preserves epimorphisms, we obtain the quotient monad, which we call $\Rquot$, 
and the epimorphic projection $\pi: \hat{\Upsilon} \to \Rquot$. 
We now equip $\Rquot$ with a $\Sigma$-action, and show that the induced model is initial,
in four steps:

\begin{enumerate}

\item
We equip  $\Rquot$ with a $\Sigma$-action, i.e., with a morphism
of $\Rquot$-modules $m_{\Rquot}:\Sigma(\Rquot)\rightarrow \Rquot$.
We define $u : \Upsilon(\hat{\Upsilon}) \rightarrow \Sigma(\Rquot)$ as $u=F_\Rquot\circ \Upsilon({\pi})$. 
% or, equivalently by naturality of $F$, as $u = b_\Rquot \circ F_R$. 
Then $u$ is epimorphic, by composition of epimorphisms and by using Corollary~\ref{t:signature-epi-morph}.
Let  $m_{\hat{\Upsilon}}:
\Upsilon(\hat{\Upsilon}) \rightarrow  \hat{\Upsilon}$ be the action of the initial model of $\Upsilon$.
We define $m_\Rquot$ as the unique morphism making the following diagram commute in the category of endofunctors on $\Set$:

\[
\begin{tikzcd}[sep=large]
\Upsilon(\hat{\Upsilon}) \arrow[r, "m_{\hat{\Upsilon}}"] \arrow[d, "u", swap]  & \hat{\Upsilon}
\arrow[d, "\pi"] 
\\
\Sigma(\Rquot) \arrow[r, swap, "m_\Rquot", dashed]  & \Rquot
\end{tikzcd}
\]

Uniqueness is given by the pointwise surjectivity of $u$. Existence follows from the compatibility of $m_{\hat{\Upsilon}}$ with the congruence $\sim_X$.
The diagram necessary to turn $m_\Rquot$ into a module morphism on $\Rquot$ is proved by pre-composing it with the epimorphism
$(\Sigma({\pi})\circ F_{\hat{\Upsilon}})\cdot \pi : \Upsilon(\hat{\Upsilon})\cdot \hat{\Upsilon}\to \Sigma(\Rquot)\cdot\Rquot$
(this is where the preservation of epimorphims by $\Upsilon(\hat{\Upsilon})$ is required) and unfolding the definitions.
\item
Now, $\pi$ can be seen as a morphism of $\Upsilon$-models between $\hat{\Upsilon}$ and $F^* \Rquot$, by naturality of $F$ and using the previous diagram.

It remains to show that $(\Rquot, m_\Rquot)$ is initial in the category of $\Sigma$-models.
\item
Given a $\Sigma$-model $(S,m_s)$, the initial morphism of $\Upsilon$-models $i_S : \hat{\Upsilon} \to F^*S$
induces a monad morphism $\iota_S:\Rquot \rightarrow S$.
We need to show that the morphism $\iota$ is a morphism of $\Sigma$-models.
Pre-composing the involved
diagram by the epimorphism $\Sigma({\pi})\circ F_{\hat{\Upsilon}} : \Upsilon(\hat\Upsilon)\to \Sigma(\Rquot)$ and unfolding
the definitions show that $\iota_S:\Rquot \rightarrow S$ is a morphism of $\Sigma$-models.

\item
We show that $\iota_S$ is the only morphism $\Rquot \rightarrow S$. Let $g$ be such a morphism. 
Then $g \circ \pi : \hat{\Upsilon}\rightarrow S$ defines a morphism in the category of $\Upsilon$-models. 
Uniqueness of $i_S$ yields $g\circ \pi = i_S$, and by uniqueness of the diagram defining $\iota_S$ it follows that $g=i'_S$.
\qedhere
\end{enumerate}
\end{proof}

\begin{lem}[\coqident{Signatures.BindingSig}{algebraic_model_Epi} and \coqident{Signatures.BindingSig}{BindingSig_on_model_isEpi}]
  \label{lem:alg-preserve-epis}
  Let $\Sigma$ be an algebraic signature.
  Then $\hat{\Sigma}$ and $\Sigma(\hat{\Sigma})$ preserve epimorphisms.
  \end{lem}
\begin{proof}
The initial model of an algebraic signature $\Sigma$ is obtained as the initial
chain of the endofunctor $R\mapsto \mathsf{Id} + \Sigma(R)$, where $\Sigma$ denotes
(by abuse of notation) the endofunctor on endofunctors on $\Set$ corresponding to the signature $\Sigma$.
Then the proof follows from the fact that this endofunctor preserves preservation of epimorphisms.
\end{proof}

\begin{proof}[Proof of Thm.~\ref{thm:pres_sig_repr}]
 Let $p : \Upsilon \to \Sigma$ be a presentation of $\Sigma$.
 We need to construct a representation for $\Sigma$.

 Since the  signature $\Upsilon$ is algebraic, it is representable
 (by Theorem~\ref{t:alg-sig-representable}),
 and it is an \epiar{} (by Example~\ref{ex:alg-epi-ar}).
 We can thus instantiate Lemma~\ref{lem:th-epi-ini} to see that $\Sigma$ is representable,
 thanks to Lemma~\ref{lem:alg-preserve-epis}.
\end{proof}

\section{Examples of presentable signatures}\label{s:examples}

Complex signatures are naturally built as the sum of basic components,
generally referred as \enquote{arities} (which in our settings are
signatures themselves, see remark after Example~\ref{ex:signatures}).
Thanks to Proposition~\ref{prop:colim-of-presentables}, direct sums (or,
more generally, colimits) of presentable signatures are presentable,
hence representable by Theorem~\ref{thm:pres_sig_repr}.

In this section, we show that, besides algebraic signatures, there are
other interesting examples of signatures which are presentable, and
which hence can be \emph{safely} added to any presentable signature.
\emph{Safely} here means that the resulting signature is still
presentable.

\subsection{Post-composition with a presentable functor}\label{sec:post-comp-with-presentable-functor}

A functor $F : \Set \to \Set$ is \emph{polynomial} if it is
of the form $FX = \coprod_{n\in \NN} a_n \times X^n$ 
for some sequence $(a_n)_{n\in\NN}$ of sets.
Note that if $F$ is polynomial, then the signature $F \cdot \Theta$ is algebraic.

\begin{defi}
Let $G : \Set \to \Set$ be a functor. A \emph{presentation of $G$} is a pair
consisting of
a polynomial functor  $F : \Set \to \Set$ and an epimorphism $p : F \to G$.
The functor $G$ is called \emph{presentable} if there is a presentation of $G$.
\end{defi}

\begin{prop}\label{prop:presentable_postcomp}
Given a presentable functor $G$, the signature $G\cdot \Theta$ is presentable.
\end{prop}
\begin{proof}
 Let $p : F \to G$ be a presentation of $G$;
 then a presentation of $G\cdot \Theta$ is given by the induced epimorphism $F\cdot \Theta \to G\cdot\Theta$.
\end{proof}

The next statement shows how the difference between finitary endofunctors and our presentable endofunctors is small:
\begin{prop}
  Here we assume the axiom of excluded middle.
 An endofunctor on $\Set$ is presentable if and only if it is finitary
 (i.e., it preserves filtered colimits).
\end{prop}
\begin{proof}
  This is a corollary of Proposition~5.2 of~\cite{Adamek:2004:tree-coalgebras},
  since $\omega$-accessible functors are exactly the finitary ones.
  \end{proof}
We now give several examples of presentable signatures obtained from presentable functors.

\subsubsection{Example: Adding a syntactic commutative binary operator, e.g., parallel-or}
\label{ex:binary-commutative}

Consider the functor $\squarefunctor : \Set \to \Set$ mapping a set $X$ to $X \times X$;
it is polynomial.
The associated signature $\squarefunctor \cdot \Theta$ encodes a binary operator,
such as the application of the lambda calculus.

Sometimes such binary operators are asked to be \emph{commutative}; a
simple example of such a commutative binary operator is the addition
of two numbers.

Another example, more specific to formal computer languages, is
a \enquote{concurrency} operator $P \mathbin{|} Q$ of a process
calculus, such as the $\pi$-calculus, for which it is natural to require
commutativity as a structural congruence relation:
$P \mathbin{|} Q \equiv Q \mathbin{|} P$.

Such a commutative binary operator can be specified via the following presentable signature:
we denote by $\mathcal{S}_2 : \Set \to \Set$ the endofunctor that assigns,
to each set $X$, the set $(X \times X)/(x,y)\sim(y,x) $ of unordered pairs of elements of $X$.
This functor is presented by the obvious projection $\squarefunctor \to \mathcal{S}_2$.
By Proposition~\ref{prop:presentable_postcomp}, the signature $\mathcal{S}_2 \cdot \Theta$
is presentable, it encodes a commutative binary operator.

\subsubsection{Example: Adding a maximum operator}

Let $\listfunctor : \Set \to \Set$ be the functor associating, to any
set $X$, the set $\listfunctor(X)$ of (finite) lists with entries in $X$;
specifically, it is given on objects as $X \mapsto \coprod_{n \in \NN} X^n$.

We now consider the syntax of a ``maximum'' operator, acting, e.g.,
on a list of natural numbers:
\[ \textsf{max} : \listfunctor(\NN) \to \NN \]
It can be specified via the algebraic signature $\listfunctor \cdot \Theta$.

However, this signature is \enquote{rough} in the sense that it does not take into account 
some semantic aspects of a maximum operator, such as invariance under repetition or permutation of elements in a list.

For a finer encoding, consider the functor $\finpower : \Set \to \Set$ associating, to a set $X$, the set $\finpower(X)$ of its finite subsets.
This functor is presented by the epimorphism $\listfunctor \to \finpower$.

By Proposition~\ref{prop:presentable_postcomp}, the signature
$\finpower\cdot\Theta$ is presentable; it encodes the syntax
of a \enquote{maximum} operator accounting for invariance under repetition or permutation of elements in a list.

\subsubsection{Example: Adding an application à la Differential LC}

Let $R$ be a commutative (semi)ring. 
To any set $S$, we can associate the \emph{free $R$-module} $R\langle S \rangle$;
its elements are formal linear combinations $\sum_{s\in S} a_ss$ of elements of $S$ with coefficients $a_s$ from $R$;
with $a_s = 0$ almost everywhere.
Ignoring the $R$-module structure on $R\langle S \rangle$,
this assignment induces a functor $R\langle \_ \rangle : \Set \to \Set$ with the obvious action on morphisms.
For simplicity, we restrict our attention to the semiring $(\NN, + , \times)$.

This functor is presentable: a presentation is given
by the polynomial functor $\listfunctor : \Set \to \Set$
and the epimorphism
\begin{align*}
           p : \listfunctor &\rar \NN \langle \_ \rangle 
           \\
               p_X \left([x_1,\ldots,x_n]\right) & := x_1 + \ldots + x_n \enspace .
\end{align*}

By Proposition~\ref{prop:presentable_postcomp}, this yields a presentable signature, which we call $\NN\langle\Theta\rangle$.

The Differential Lambda Calculus (DLC) of Ehrhard and Regnier~\cite{DBLP:journals/tcs/EhrhardR03} is a lambda calculus with operations suitable to express differential constructions.
The calculus is parametrized by a semiring $R$; again we restrict to $R = (\NN, + , \times)$.

DLC has a binary \enquote{application} operator, written $(s)t$, where $s \in T$ is an element of the inductively defined set $T$ of terms and 
$t \in \NN\langle T\rangle$ is an element of the free $(\NN, + , \times)$-module.
This operator is thus specified by the presentable signature  $\Theta \times \NN\langle\Theta\rangle$.

\subsection{Example: Adding a syntactic closure operator}
\label{sec:syntactic-closure}

Given a quantification construction (e.g., abstraction, universal or
existential quantification), it is often useful to take the associated
closure operation.  One well-known example is the universal closure of
a logic formula. Such a closure is 
invariant under permutation of the fresh variables.
A closure can be syntactically encoded in a rough way by iterating the
closure with respect to one variable at a time.
Here our framework allows a refined syntactic encoding which we
explain below.

Let us start with binding a fixed number $k$ of fresh variables.
The elementary signature $\Theta^{(k)}$ already
specifies an operation that binds $k$ variables. However, this encoding does not reflect invariance under variable permutation.
To enforce this invariance, it suffices to quotient the signature $\Theta^{(k)}$ with respect to the
action of the group $S_k$ of permutations of the set $k$,
that is, to consider the colimit
of the following one-object diagram:
\[
\begin{tikzcd}[sep=large]
\Theta^{(k)}\ar[loop above]{}{\Theta^{(\sigma)}}
\end{tikzcd}
\]
where $\sigma$ ranges over the elements of $S_k$.
We denote by $\symdsig{k}$ the resulting signature presented by the projection $\Theta^{(k)} \to \symdsig{k}$.
By universal property of the quotient, a model of it consists of a monad $R$ with
an action $m : R^{(k)} \rightarrow R$ that satisfies the required invariance.

Now, we want to specify an operation which binds an arbitrary number of fresh variables, as expected
from a closure operator. One rough solution is to consider the coproduct
$\coprod_{k} \symdsig{k}$. However, we encounter a similar inconvenience as for $\Theta^{(k)}$.
Indeed, for each $k' > k$, each term already encoded by the signature
$\symdsig{k}$ may be considered again, encoded (differently)
through $\symdsig{k'}$.

Fortunately, a finer encoding is provided by the following simple colimit of presentable signatures. 
The crucial point here is that, for each $k$, all natural injections from $\Theta^{(k)}$  to $\Theta^{(k+1)}$ induce the same 
canonical injection from $\symdsig{k}$ to $\symdsig{k+1}$. 
We thus have a natural colimit for the sequence $k \mapsto \symdsig{k}$ and thus a signature 
$\colim_{k}  \symdsig{k}$ which, as a colimit of presentable signatures, is presentable
(Proposition~\ref{prop:colim-of-presentables}).

Accordingly, we define a total closure on a monad $R$ to be an action of the signature 
$\colim_{k}  \symdsig{k}$ in $R$.
It can easily be checked that a model of this signature is a monad $R$
together with a family of module morphisms
$(e_k:R^{(k)} \rightarrow R)_{k\in \NN}$
  compatible in the sense that
  for each injection $i: k\rightarrow k'$ the following diagram commutes:
\[
 \begin{xy} 
   \xymatrix@C=5em@R=4em{ 
                     R^{(k)}  \ar[rd]_{e_k}   \ar[r]^-{R^{(i)}}&   **[r] R^{(k')} \ar[d]^{e_{k'}} 
                     \\       
                                & R 
   }  
 \end{xy} 
\]

\subsection{Example: Adding an explicit substitution}
\label{sec:kleisli_substitution}

\emph{Explicit substitution} was introduced by Abadi et
al.~\cite{Abadi:1989:ES:96709.96712} as a theoretical device to study
the theory of substitution and to describe concrete implementations of
substitution algorithms.  In this section, we explain how we can
extend any presentable signature with an explicit substitution
construction, and we offer some refinements from a purely syntactic
point of view.  In fact, we will show three solutions, differing in the
amount of \enquote{coherence} which is handled at the syntactic level
(e.g.,~invariance under permutation and weakening).  We follow the
approach initiated by Ghani, Uustalu, and Hamana in
\cite{DBLP:journals/lisp/GhaniUH06}.
With respect to their work, one key difference is that our approach does not require the notion of strength. 

 Let $R$ be a monad.
We have already considered (see Lemma~\ref{l:linear-sigma}) the (unary) substitution $\sigma_R : R'\times R \rightarrow R$.
More generally, we have the sequence of substitution operations
\begin{equation}
  \label{eq:subst}
\subst_p : R^{(p)} \times R^{p} \rar R.
\end{equation}
We say that  $\subst_p$ is the $p$-substitution in $R$; it simultaneously replaces the $p$ extra variables in its first argument with 
the $p$ other arguments, respectively.  (Note that $\subst_1$ is the original $ \sigma_R$). 

We observe that, for fixed $p$, the group $S_p$ of permutations on $p$ elements has a natural action on
$R^{(p)} \times R^{p}$, and that $\subst_p$ is invariant under this action.

Thus, if we fix an integer $p$, there are two ways to internalise $\subst_p$ in the syntax:
we can choose the elementary signature 
$\Theta^{(p)}\times\Theta^p$, which is rough in the sense that the above invariance is
not reflected; 
and, alternatively, if we want  to reflect the permutation invariance syntactically, we can choose the quotient $Q_p$ of the above signature by the action 
of $S_p$.

By universal property of the quotient, a model of our quotient $Q_p$ is given by a monad
$R$ with an action $m : R^{(p)}\times R^p \to R$ satisfying the desired invariance.

Before turning to the encoding of the entire series $(\subst_p)_{p\in \NN}$, we recall how, as noticed already in \cite{DBLP:journals/lisp/GhaniUH06}, this series enjoys further coherence. 
In order to explain this coherence, we start with two natural numbers $p$ and $q$ and the module  $R^{(p)}\times R^q$. Pairs in this module are almost ready for substitution: 
what is missing is a map   $u: p \rar q$ between the corresponding standard finite sets.
But such a map can be used in two ways: letting $u$ act covariantly on the first factor leads us into
$R^{(q)}\times R^q$ where we can apply $\subst_q$; while letting $u$ act contravariantly on the second factor leads us into
$R^{(p)}\times R^p$ where we can apply $\subst_p$. The good news is that we obtain the same result. More precisely, the following diagram is commutative:
\begin{equation}\label{eq:kleisli-commute}
  \begin{gathered}
\begin{xy}
\xymatrix@C=2cm@R=1.4cm{
 **[l] R^{(p)} \times R^{q}
 \ar[r]^{R^{(p)} \times R^u}
 \ar[d]_{R^{(u)} \times R^{q}}
 &
  **[r] R^{(p)} \times R^p   
 \ar[d]^{\subst_p}
 \\
 **[l] R^{(q)} \times R^{q}   
 \ar[r]_{\subst_q}
 &  
 **[r] R
 }
\end{xy}
  \end{gathered}
\end{equation}
Note that in the case where $p$ equals $q$ and $u$ is a permutation, we recover exactly the invariance by permutation considered earlier.

 Abstracting over the numbers $p, q$ and the map $u$, this exactly means that our series factors through
the coend
  $ \coend{R}$,
where covariant (resp.~contravariant)
occurrences of the bifunctor have been underlined (resp.~overlined), and
the category $\NN$ is the full subcategory of $\Set$
whose objects are natural numbers. Thus we have a canonical morphism
  \[ \intsubst_R : \coend{R} \rar R. \]
Abstracting over $R$, we obtain the following:

\begin{defi}
  \label{def:usubst}
  The \emph{integrated substitution}
  \[ \intsubst : \coend{\Theta} \rar \Theta \] is the signature
  morphism obtained by abstracting over $R$ the linear morphisms
  $\intsubst_R$.
\end{defi}

Thus, if we want to internalise the whole sequence $(\subst_p)_ {p: \NN} $ in the syntax, we have at least three solutions: we can choose the algebraic signature 
\[ \coprod_{p: \NN}  \Theta^{(p)}\times\Theta^p \] which is rough in the sense that the above invariance and coherence is not reflected; 
we can choose the presentable signature 
\[\coprod_{p: \NN}  Q_p, \] 
which reflects the invariance by permutation, but not more; and finally, if we want to reflect the whole coherence syntactically, we can choose the presentable signature  
\[\coend{\Theta} . \]

Thus, whenever we have a presentable signature, we can safely extend it
by adding one or the other of the three above signatures, for a (more or less coherent) explicit substitution.

\subsection{Example: Adding a coherent fixed-point operator}

In the same spirit as in the previous section, we define, in this section,
\begin{itemize}
\item for each $n \in \NN$, a notion of \emph{$n$-ary fixed-point operator} in
  a monad;
\item a notion of \emph{coherent fixed-point operator} in a monad, which
  assigns, in a \enquote{coherent} way, to each $n \in \NN$, an $n$-ary fixed-point operator.
\end{itemize}
We furthermore explain how to safely extend any syntax generated by a presentable signature with a
syntactic coherent fixed-point operator.

There is one fundamental difference between the integrated
substitution of the previous section and our coherent fixed points: while
every monad has a canonical integrated substitution, this is not the
case for coherent fixed-point operators.

Let us start with the unary case.
\begin{defi}\label{def:unary-fp-op}
  A \emph{unary fixed-point operator for a monad $R$} is a module
  morphism $f$ from $R'$ to $R$ that makes the following diagram
  commute,
  \[
    \begin{tikzcd}[sep=large]
      R'
      \arrow[rr, "(id_{R'} \text{,} f)"]
      \arrow[rd,"f",swap]
      & & R' \times R \arrow[ld, "\sigma"] \\
      & R
    \end{tikzcd}
  \]
  where $\sigma$ is the substitution morphism defined in
  Lemma~\ref{l:linear-sigma}.
\end{defi}

Accordingly, the signature for a syntactic unary fixpoint operator is
$\Theta '$, ignoring the commutation requirement (which we plan to address
in a future work by extending our framework with equations).

Let us digress here and examine what the unary fixpoint
operators are for the lambda calculus, more precisely, for the monad
$\LCb$ of the lambda calculus modulo $\beta$- and $\eta$-equivalence.
How can we relate the above notion to the
classical notion of fixed-point combinator?  Terms are built out of
two constructions, $\app:\LCb \times \LCb \rightarrow \LCb$ and
$\abs:\LCb'\rightarrow \LCb$.  A fixed-point combinator is a term $Y$
satisfying, for any (possibly open) term $t$, the equation
\begin{equation*}
  \app(t, \app(Y, t)) = \app(Y,t).
\end{equation*}
Given such a combinator $Y$, we define a module morphism
$\hat{Y}:\LCb'\rightarrow \LCb$. It associates, to any term $t$
depending on an additional variable $*$, the term
$\hat Y (t) := \app(Y,\abs \: \, t)$.  This term satisfies
$t[\hat Y (t)/*] = \hat Y (t)$, which is precisely the commutativity of the diagram of
Definition~\ref{def:unary-fp-op} that $\hat Y$ must satisfy to be a
unary fixed-point operator for the monad $\LCb$.  Conversely, we have:
\begin{prop}\label{thm:fixed-pt-combinator-operator}
  Any fixed-point combinator in $\LCb$ comes from a unique fixed-point operator.
\end{prop}
\begin{proof}
  We construct a bijection between 
  the set $\LCb(0)$ of closed terms on the one hand 
  and the set of module morphisms
  from $\LCb'$ to $\LCb$ satisfying the fixed-point property on the other hand.
  
  A closed lambda term $t$ is mapped to the morphism
  $u\mapsto \hat t\: u := \app(t,\abs\:\,u)$. We have already seen that
  if $t$ is a fixed-point combinator, then $\hat t$ is a fixed-point operator.

  For the inverse function, note that
  a module morphism $f$ from $\LCb'$ to $\LCb$
  induces a closed term $Y_f := \abs (f_1(\app(*,**)))$ where
  $f_1:\LCb(\{*,**\})\rightarrow \LCb(\{*\})$.

A small calculation shows that $Y\mapsto \hat Y$ and $f\mapsto Y_f$ are inverse to each other.

It remains to be proved that if $f$ is a fixed-point operator, then $Y_f$
  satisfies the fixed-point combinator equation. Let $t\in \LCb X$, then we have
\begin{align}
                           \app(Y_f,t) & = \app(\abs \: f_1(\app(*,**)),t)
                                         \label{eq:fix8}
                                         \\ & 
                                              = f_{X}(\app(t,**)) \label{eq:fix3}
                                         \\ & 
                                              = \app(t,\app(Y_f,t))
                                              \label{eq:fix4}
  \end{align}
  where \eqref{eq:fix3} comes from the definition of a fixed-point operator.
  Equality \eqref{eq:fix4} follows from the equality
  $\app(Y_f,t) = f_{X}(\app(t,**))$, which is obtained by chaining the equalities from \eqref{eq:fix8} to \eqref{eq:fix3}.
  This concludes the construction of the bijection.
\end{proof}

After this digression, we now turn to the $n$-ary case.
\begin{defi}
\hfill
\begin{enumerate}
\item
  A \emph{rough $n$-ary fixed-point operator} for a monad $R$ is a
  module morphism $f: (R^{(n)})^n \to R^n$ making the following
  diagram commute: 
  \[
    \xymatrix@C=3cm@R=1.2cm{
      (R^{(n)})^n
      \ar[r]^-{id_{(R^{(n)})^n}\text{,} f\text{,}..\text{,}f}
      \ar[d]_f
      &
      (R^{(n)})^n
      \times
      (R^{n})^n
      \ar[d]^{\cong}
      % \ar@{}[d]|*=0[@]{\cong}
      \\
       R^n
       &
      (R^{(n)}\times R^n)^n
       \ar[l]^{(\subst_n)^n}
       }
  \]
  where $\subst_n$ is the $n$-substitution as in
  Section~\ref{sec:kleisli_substitution}.
\item
An \emph{$n$-ary fixed-point operator} is just a rough $n$-ary fixed-point operator which is furthermore invariant under the natural action of the permutation group $S_n$.
\end{enumerate}
\end{defi}

\noindent
The type of $f$ above is canonically isomorphic to
\begin{equation*} 
  (R^{(n)})^n + (R^{(n)})^n + \ldots + (R^{(n)})^n \to R, 
\end{equation*} 
which we abbreviate to
$n \times (R^{(n)})^n \to R$. 

Accordingly, a natural signature for encoding a syntactic rough $n$-ary
fixpoint operator is $n \times (\Theta^{(n)})^n $.  

Similarly, a natural signature for encoding a syntactic $n$-ary fixpoint
operator is $(n \times (\Theta^{(n)})^n )/S_n$
obtained by quotienting the previous signature by the action of $S_n$.

Now we let $n$ vary
and say that a \emph{total fixed-point operator} on a given
monad $R$ assigns to each $n \in \NN$ an $n$-ary fixpoint operator on
$R$. Obviously, the natural signature for the encoding of a syntactic
total fixed-point operator is $\coprod_n (\Theta^{(n)})^n /S_n$.
Alternatively, we may wish to discard those total fixed-point
operators that do not satisfy some coherence conditions analogous to
what we encountered in Section~\ref{sec:kleisli_substitution}, which we now introduce.

Let $R$ be a monad with a sequence of module morphisms
$\fix_n : n \times (R^{(n)})^n \to R$.  We call this family
\emph{coherent} if, for any $p, q \in \NN$ and $u : p \to q$, the following diagram commutes:

\begin{equation}\label{eq:fix-commute}
  \begin{xy}
    \xymatrix@C=2cm@R=1.2cm{
      **[l] p \times (R^{(p)})^q
      \ar[r]^{p \times (R^{(p)})^u}
      \ar[d]_{u \times (R^{(u)})^q}
      &
      **[r] p \times (R^{(p)})^p
      \ar[d]^{\mathsf{fix}_{p}}
      \\
      **[l] q \times (R^{(q)})^q
      \ar[r]_{\mathsf{fix}_q}
      &
      **[r] R
    }
  \end{xy}
\end{equation}
These conditions have an interpretation in terms of a coend, just as
we already encountered in Section~\ref{sec:kleisli_substitution}. This leads us to the following
\begin{defi}
  \label{def:coherent-seq-of-fixed}
  Given a monad $R$, we define a \emph{coherent fixed-point operator on
    $R$} to be a module morphism from $\fixcoend{R}$ to $R$ where, for
  every $n\in\NN$, the $n$-th component is a (rough)\footnote{As in Section \ref{sec:kleisli_substitution},
    the invariance follows from the coherence.} $n$-ary fixpoint
  operator.
\end{defi}

Thus, given a presentable signature $\Sigma$, we can safely extend it
with a syntactic coherent fixed-point operator by adding the presentable
signature 
\[  \fixcoend{\Theta}\] 
to $\Sigma$.

\section{Recursion}
\label{sec:recursion}

Initiality can be seen as an abstract way to present recursion.
In this section, we collect several examples of recursive maps that can be derived from initiality.

\subsection{Example: Translation of intuitionistic logic into linear logic}
We start with an elementary example of translation of syntaxes using initiality, namely the translation
of second-order intuitionistic logic into second-order linear logic 
\cite[page 6]{Girard:LL}.
The syntax of second-order intuitionistic logic can be defined with
one unary operator $\neg$,
three binary operators $\lor$,  $\land$ and $\Rightarrow$, and two binding operators
$\forall$ and $\exists$. The associated (algebraic) signature is
$\Sigma_{LJ} = \Theta + 3 \times \Theta^2 + 2 \times \Theta'$.
As for linear logic, there are four constants $\top,\bot,0,1$,
two unary operators $!$ and $?$, five binary operators
$\&$,
$\invamp$,
$\otimes$,
$\oplus$,
$\multimap$
and two binding operators $\forall$ and $\exists$.
The associated (algebraic) signature is
$\Sigma_{LL}
= 4 \times *
+ 2 \times \Theta
+ 5 \times \Theta^2
+ 2 \times \Theta'$.

By universality of the coproduct, a model of $\Sigma_{LJ}$ is given by a monad $R$ with module morphisms:
\begin{align*}
r_\neg &: R \rar R
\\
r_\land,r_\lor,r_\Rightarrow &: R\times R \rar R
\\
  r_\forall,r_\exists &: R' \rar R
\end{align*}
and similarly, we can decompose an action of $\Sigma_{LL}$ into as many components as there are operators.

The translation will be a morphism of monads between the initial models
(i.e.~the syntaxes)
$o:\hat \Sigma_{LJ} \rar \hat \Sigma_{LL}$
coming from 
the initiality of $\hat \Sigma_{LJ}$. Indeed, %
equipping $\hat \Sigma_{LL}$ with an action $r'_\alpha:\alpha(\hat \Sigma_{LL})\rar\hat \Sigma_{LL}$
for each operator $\alpha$ of intuitionistic logic
($\neg$,$\lor$,$\land$,$\Rightarrow$,$\forall$  and $\exists$)
  yields a morphism of monads 
$o:\hat \Sigma_{LJ} \rar \hat \Sigma_{LL}$ such that $o(r_\alpha(t))=r'_\alpha(\alpha(o)(t))$ for each $\alpha$.

  The definition of $r'_\alpha$ is then straightforward to devise, following the standard translation given
  on the right:
\begin{align*}
  r'_\neg  & = r_\multimap \circ (r_! \times r_0)
  & (\neg A)^o := (!A)\multimap 0
  \\
  r'_\land & = r_\&
  & (A \land B)^o := A^o \& B^o
  \\
  r'_\lor  & = r_\oplus \circ( r_! \times r_!)
  &
  (A \lor B)^o := !A^o\oplus!B^o
  \\
  r'_\Rightarrow & = r_\multimap \circ (r_! \times id)
  & (A \Rightarrow B)^o := !A^o \multimap B^o
  \\
  r'_\exists &  =  r_\exists\circ r_!
  & (\exists x A)^o := \exists x ! A^o
  \\
  r'_\forall &  =  r_\forall
  & (\forall x A)^o := \forall x  A^o
\end{align*}

The induced action of $\Sigma_{LJ}$ in the monad $\hat \Sigma_{LL}$ yields the desired translation morphism
$o:\hat \Sigma_{LJ} \rightarrow \hat \Sigma_{LL}$. Note that variables are automatically preserved by the translation because $o$ is a monad morphism.

\subsection{Example: Computing the set of free variables}
\label{sec:iter-free}

As above, we denote by $\PP X$ the powerset of $X$.  The union gives us a
composition operator $\PP(\PP X) \to \PP X$ defined by
$u \mapsto \bigcup_{s\in u} s$, which yields a monad
structure on $\PP$.

We now define an action of the signature of lambda calculus
$\Sigma_\LC$ in the monad $\PP$.  We take the binary union operator
$\cup\colon \PP \times \PP \to \PP$ as action of the application signature
$\Theta\times\Theta$ in $\PP$; this is a module morphism since binary union distributes over union of sets.
Next, given
$S \in \PP(X+*)$ we define $\maybeinv_X(S) = S \cap X$.  
This defines a morphism of modules 
$\maybeinv \colon \PP' \to \PP$;
a small calculation using a distributivity law of binary intersection over union of sets shows that 
this natural transformation is indeed linear.
It can hence be used to model the abstraction signature $\Theta'$ in $\PP$.

Associated to this model of $\Sigma_\LC$ in $\PP$ we have an initial
morphism $\mathsf{free}\colon \LC \to \PP$.  
Then, for any $t \in \LC(X)$, the set $\mathsf{free}(t)$ is the set of
free variables occurring in $t$.

\subsection{Example: Computing the size of a term}
\label{sec:iter-size}

We now consider the problem of computing the \enquote{size} of a
$\lambda$-term, that is, for any set $X$, a function
  $s_X \colon \LC(X) \rar \NN$
such that
\begin{align*}
  s_X(x) &= 0\qquad\text{($x \in X$ variable)}\\
  s_X(\abs(t))  &= 1 + s_{X+*} (t)\\
  s_X(\app(t,u)) &= 1 + s_X(t) + s_X(u) 
\end{align*}
To express this map as a morphism of models, we first 
need to find a suitable monad underlying the target model.
The first candidate, the constant functor $X \mapsto \NN$, does not admit
a monad structure; the problem lies in finding a suitable unit for the monad.
(More generally, given a monad $R$ and a set $A$, the functor $X \mapsto R(X) \times A$
does not admit a monad structure whenever $A$ is not a singleton.)

This problem hints at a different approach to the original question:
instead of computing the size of a term (which
is $0$ for a variable), we compute a generalized size $gs$ which
depends on arbitrary (formal) sizes attributed to variables. We have
\begin{equation*}
  gs: \prod_{X:\Set} \Big(\LC(X) \to (X \to \NN) \to \NN\Big)
\end{equation*}
Here, unsurprisingly, we recognize the continuation monad (see also
\cite{DBLP:conf/tlca/JohannG07} for the use of continuation for
implementing complicated recursion schemes using initiality)
\begin{equation*}
  \Cont_\NN := X \mapsto (X \to \NN) \to \NN
\end{equation*}
with multiplication $\lambda f.\lambda g.f (\lambda h.h(g))$. 

Now we can define $gs$ through initiality by endowing the monad
$\Cont_\NN$ with a structure of $\Sigma_\LC$-model as follows.

The function
\begin{math}
  \alpha(m,n) = 1 + m + n
\end{math}
induces a natural transformation
\begin{equation*}
  c_\app \colon \Cont_\NN \times \Cont_\NN \rar \Cont_\NN
\end{equation*}
and thus an action for the application signature
$\Theta\times\Theta$ in the monad $\Cont_\NN$.

Next, given a set $X$ and $k:X \to \NN$, define $\hat{k} : X+\{*\}\to \NN$ by
$\hat{k}(x)=k(x)$ for all $x\in X$ and $\hat{k}(*)=0$.
This induces a function
\begin{equation*}
\begin{array}{rcl}
  c_\abs(X) \colon  \Cont'_\NN(X)
 & \rar & \Cont_\NN(X)  \\
    t & \mapsto & (k\mapsto 1 + t( \hat{k}))
\end{array}
\end{equation*}
which is the desired action of the abstraction signature
$\Theta'$.

Altogether, the transformations $c_\app$ and $c_\abs$ form the desired action of $\Sigma_\LC$ in
$\Cont_\NN$ and thus give an initial morphism, i.e., a natural
transformation $\iota\colon\LC\to \Cont_\NN$ which respects the
$\Sigma_\LC$-model structure.  Now let $0_X$ be the function that
is constantly zero on $X$.  Then the sought \enquote{size} map $ s : \prod_{X:\Set} \LC(X) \to \NN$
is given by $s_X(t) = \iota_X (t, 0_X)$.

\subsection{Example: Counting the number of redexes}
\label{sec:recur-redexes}

We now consider a function $r$
such that $r(t)$ is the number of redexes of the $\lambda$-term $t$ of
$\LC(X)$.  Informally, the equations defining $r$ are
\begin{align*}
  r(x) &= 0,\qquad\text{($x$ variable)}\\
  r(\abs(t)) &= r(t),\\
  r(\app(t,u)) &= r(t) + r(u) +
  \begin{cases}
    1 & \text{if $t$ is an abstraction} \\
    0 & \text{otherwise}
  \end{cases}
\end{align*}
In order to compute recursively the number of $\beta$-redexes in a term,
we need to keep track, not only of the number of
redexes in subterms, but also whether the head construction of subterms is
the abstraction; in the affirmative case we use the value $1$, and otherwise we use $0$.
Hence, we define a $\Sigma_\LC$-action on the monad
$W := \Cont_{\NN \times \{0,1\}}$.  We denote by $\pi_1$, $\pi_2$ the
projections that access the two components of the product
$\NN \times \{0,1\}$.

 For any set $X$ and function $k:X\to \NN\times \{0,1\}$, let us denote by
$\hat k:X+\{*\}\to \NN\times \{0,1\}$  the function which sends $x\in X$ to $k(x)$ and $*$ to $(0, 0)$.
Now, given a set $X$, consider the function
\begin{equation*}
\begin{array}{rcl}
  c_\abs(X) \colon  W'(X)
 & \rar & W(X)  \\
    t & \mapsto & (k\mapsto (\pi_1 (t(\hat k)), 1)).
\end{array}
\end{equation*}
Then $c_\abs$ is an action of the abstraction signature $\Theta'$ in $W$.

Next, we specify an action $c_\app : W \times W \to W$ of the application signature $\Theta \times  \Theta$:
Given a set $X$, consider the function
\begin{equation*}
\begin{array}{rcl}
  c_\app(X) \colon  W(X) \times W(X)
 & \rar & W(X)  \\
    (t,u) & \mapsto & (k\mapsto (\pi_1(t(k)) + \pi_1(u(k)) + \pi_2(t(k)) , 0) ).
\end{array}
\end{equation*}
Then $c_\app$ is an action of the abstraction signature $\Theta\times\Theta$ in $W$.

Overall we have a $\Sigma_\LC$-action from which we get an initial
morphism $\iota\colon \LC \to W$.  If $0_X$ is the constant function
$X\to {\NN\times\{0,1\}}$ returning the pair $(0,0)$, then
$\pi_1(\iota(0_X)) : \LC(X) \to \NN$ is the desired function $r$.

\section{Conclusion}

\subsection*{Summary}
We have presented notions of \emph{signature} and \emph{model of a signature}.
A \emph{representation of a signature} is an initial object in its category of models---a \emph{syntax}.
We have defined a class of \emph{presentable} signatures, which contains traditional algebraic signatures,
and which is closed under various operations, including colimits.
One of our main results says that any presentable signature is representable.

One difference to other work on Initial Semantics (cf.\ Section~\ref{ss:related})
is that we do not rely on the notion of strength.
However, a signature endofunctor with strength as used in the aforementioned articles 
can be translated to a signature as presented in this work (Proposition~\ref{prop:sigs_w_strength}).

\subsection*{Acknowledgements}
We thank the anonymous referees and the editor for their helpful and constructive comments.

Ahrens acknowledges the support of the Centre for Advanced Study (CAS) in Oslo, Norway, which funded and hosted the research project \emph{Homotopy Type Theory and Univalent Foundations} during the 2018/19 academic year.

Maggesi acknowledges the support of Gruppo Nazionale per le Struttore Algebriche, Geometriche e le loro Applicazioni (GNSAGA), Istituto Nazionale di Alta Matematematica ``F.~Severi'' (INdAM), and Italian Ministry of Education, University and Research (MIUR).

This work has partly been funded by the CoqHoTT ERC Grant 637339 and by the EPSRC Grant EP/T000252/1.
This material is based upon work supported by the Air Force Office of Scientific Research under award number FA9550-17-1-0363.

\bibliographystyle{alpha}
\bibliography{strengthened}

\end{document}